\documentclass[12pt,preprint]{emulateapj}
\usepackage{graphicx, color}
\slugcomment{Submitted 2016 February 6, Accepted 2016 May 23}

\shorttitle{X-ray and optical studies of type I Seyfert NGC 3516}
\shortauthors{Noda et al.}

\begin{document}


\title{X-ray and Optical Correlation of Type I Seyfert NGC 3516 \\ Studied with  \textit{Suzaku} 
and Japanese Ground-Based Telescopes \\}


\author{Hirofumi Noda\altaffilmark{1,2}, Takeo Minezaki\altaffilmark{3}, 
Makoto Watanabe\altaffilmark{4,14}, 
Mitsuru Kokubo\altaffilmark{3,5},
Kenji Kawaguchi\altaffilmark{6}, \\Ryosuke Itoh\altaffilmark{6},
Kumiko Morihana\altaffilmark{7}, Yoshihiko Saito\altaffilmark{8}, 
Hikaru Nakao\altaffilmark{4}, Masataka Imai\altaffilmark{4}, 
Yuki Moritani\altaffilmark{9}, Katsutoshi Takaki\altaffilmark{6}, Miho Kawabata\altaffilmark{6},
Tatsuya Nakaoka\altaffilmark{6}, Makoto Uemura\altaffilmark{10}, 
Koji Kawabata\altaffilmark{10}, \\Michitoshi Yoshida\altaffilmark{10}, 
Akira Arai\altaffilmark{7, 15}, Yuhei Takagi\altaffilmark{7}, 
Tomoki Morokuma\altaffilmark{3}, 
Mamoru Doi\altaffilmark{3} , Yoichi Itoh\altaffilmark{7}, \\
Shin'ya Yamada\altaffilmark{11}, Kazuhiro Nakazawa\altaffilmark{12}, 
Yasushi Fukazawa\altaffilmark{6}, and Kazuo Makishima\altaffilmark{12, 13}}

\altaffiltext{1}{Frontier Research Institute for Interdisciplinary Sciences, Tohoku University,  6-3 Aramakiazaaoba, Aoba-ku, Sendai, Miyagi 980-8578, Japan, e-mail: hirofumi.noda@astr.tohoku.ac.jp}
\altaffiltext{2}{Astronomical Institute, Tohoku University, 6-3 Aramakiazaaoba, Aoba-ku, Sendai, Miyagi 980-8578, Japan}
\altaffiltext{3}{Institute of Astronomy, School of Science, The University of Tokyo, Mitaka, Tokyo 181-0015, Japan}
\altaffiltext{4}{Department of Cosmosciences, Hokkaido University, Kita 10, Nishi 8, Kita-ku, Sapporo, Hokkaido 060-0810, Japan}
\altaffiltext{5}{Department of Astronomy, School of Science, the University of Tokyo,
7-3-1 Hongo, Bunkyo-ku, Tokyo 113-0033, Japan}
\altaffiltext{6}{Department of Physical Science, Hiroshima University, 1-3-1 Kagamiyama, Higashi-Hiroshima,
Hiroshima 739-8526, Japan}
\altaffiltext{7}{Nishi-harima Astronomical Observatory, Center for Astronomy, University of Hyogo, 407-2 Nichigaichi, Sayo-cho, Sayo, Hyogo 670-5313, Japan}
\altaffiltext{8}{Department of Physics, Tokyo Institute of Technology, 2-12-1 Ookayama, Meguro-ku, Tokyo 152-8551, Japan}
\altaffiltext{9}{Kavli Institute for the Physics and Mathematics of the Universe
(Kavli IPMU), The University of Tokyo, 5-1-5 Kashiwa-no-Ha, Kashiwa 277-8583, Japan}
\altaffiltext{10}{Hiroshima Astrophysical Science Center, Hiroshima University,
Higashi-Hiroshima, Hiroshima 739-8526, Japan}
\altaffiltext{11}{Department of Physics, Tokyo Metropolitan University, 1-1 Minami-Osawa, Hachioji, Tokyo 192-0397 Japan}
\altaffiltext{12}{Department of Physics, School of Science, The University of Tokyo, 7-3-1 Hongo, Bunkyo-ku, Tokyo 113-0033, Japan}
\altaffiltext{13}{MAXI Team, Global Research Cluster, RIKEN, Wako, Saitama 351-0198, Japan}
\altaffiltext{14}{Department of Applid Physics, Okayama University of Science,
1-1, Ridai-cho, Kita-ku, Okayama 700-0005, Japan}
\altaffiltext{15}{Koyama Astronomical Observatory, Kyoto Sangyo University, Motoyama, Kamigamo, Kita-ku, Kyoto 603-8555, Japan }


%----------------------------------------abstract--------------------------------------------------------
\begin{abstract}
From 2013 April to 2014 April, we performed an X-ray and optical simultaneous 
monitoring of the type 1.5 Seyfert galaxy NGC~3516. It employed \textit{Suzaku},  
and 5 Japanese ground-based telescopes, the Pirka, Kiso Schmidt, Nayuta, 
MITSuME, and the Kanata telescopes.  
The \textit{Suzaku} observations were conducted seven times 
with various intervals ranging from days, weeks, to months, 
with an exposure of $\sim50$~ksec each.  
The optical $B$-band observations not only covered 
those of \textit{Suzaku} almost simultaneously, 
but also followed the source as frequently as possible. 
As a result, NGC~3516 was found in its faint phase 
with the 2--10 keV flux of $0.21$--$2.70\times10^{-11}$~erg~s$^{-1}$~cm$^{-2}$. 
The 2--45 keV X-ray spectra were composed of a dominant variable hard power-law continuum 
with a photon index of $\sim1.7$, and a non-relativistic reflection component 
with a prominent Fe-K$\alpha$ emission line.  
Producing the $B$-band light curve by differential image photometry, 
we found that the $B$-band flux changed by 
$\sim2.7\times10^{-11}$~erg~s$^{-1}$~cm$^{-2}$, which is comparable to the X-ray variation, 
and detected a significant flux correlation between the hard power-law 
component in X-rays and the $B$-band radiation, for the first time in NGC~3516.  
By examining their correlation, we found that the X-ray flux preceded that of 
$B$ band by $2.0^{+0.7}_{-0.6}$~days ($1\sigma$ error). 
Although this result supports the X-ray reprocessing model, 
the derived lag is too large to be explained by the standard view 
which assumes a ``lamppost''-type X-ray illuminator located near a standard accretion disk.  
Our results are better explained by assuming a hot accretion 
flow and a truncated disk. 

\end{abstract}
%--------------------------------------------------------------------------------------------------------------


\keywords{galaxies: active -- galaxies: individual (NGC 3516) -- galaxies: Seyfert -- X-rays: galaxies}



%--------------------Introduction---------------------
\section{Introduction}
%---------------------------------------------------------

An Active Galactic Nucleus (AGN) is known to generate multi-wavelength radiation 
by mass accretion onto a Super Massive Black Hole (SMBH) in its central engine. 
An important element of the central engine is the standard accretion disk 
(e.g., Shakura \& Suynaev 1973; Balbus \& Hawley 1991; Machida 2000),
which is expected to form at the Eddington ratio of $L_{\rm bol}/L_{\rm Edd} \sim0.1$--1 
(e.g., Abramowicz et al. 1995), and converts appreciable fraction of 
the gravitational energy release of the accreting matter into optically-thick radiation.
A part of the radiation has been observed as 
``big blue bump" seen in optical spectra from AGNs 
that have disk radiation stronger than jet emission, 
like Seyfert galaxies and quasars
(e.g., Elvis et al. 1994). 

The central engine requires another element that boosts up 
these low-energy photons into the observed X-ray signals. 
This is usually attributed to some sort of hot Maxwellian electrons, 
or ``corona'', which Compton-upscatter the seed photons into 
broadband X-ray photons. 
However, the configuration of such corona is still under big debates, 
in contrast to the well-understood standard accretion disk. 
Some X-ray studies led to an argument that a compact corona is located on a rotation axis of 
the accretion disk like a ``lamppost'', producing an extremely relativistically-broadened 
reflection component by illuminating the inner accretion disk
 (e.g., Miniutti \& Fabian 2004; Dauser et al. 2013). 
Others argue that an extended corona is present at the inner edge of an accretion disk, 
sometimes affected by partially-covering absorptions of ionized matters
(e.g., Miller et al. 2008; Miyakawa et al. 2012). 
To settle the scenario of the Comptonizing corona, 
we need to focus on primary X-ray spectra and their flux variability.

According to X-ray studies of AGNs, 
several kinds of primary X-ray signals with distinct spectral and timing properties were reported. 
A flat primary spectrum, which is reproduced by a single Power-Law (PL) 
continuum with the photon index of $ \Gamma \lesssim 1.7$, 
dominates in X-rays from relatively low Eddington-ratio AGNs (e.g., Terashima et al. 2002).   
On the other hand, a steep primary continuum, 
which can be explained by a PL model with $\Gamma \gtrsim 2.3$, 
is dominant in X-rays from highly-accreting AGNs 
like narrow-line type 1 Seyferts (e.g.,  Laor et al. 1994; Boller et al. 1996). 
Recently, Noda et al. (2011a, 2013a, and 2014) revealed that 
both these flat and steep primary continua are simultaneously present 
in the X-ray emission from multiple Seyfert galaxies. 
The presence of these different primary X-rays may represent 
presence of several distinct types of coronae with different electron 
temperatures and optical depths.
Furthermore, in order to explain a soft X-ray excess structure 
at a low energy band below $\sim 3$ keV, a soft thermal Comtonization 
continuum has been suggested, invoking yet another corona
(e.g., Mehdipour et al. 2011; Noda et al. 2011b, 2013b; Jin et al. 2013). 
The reality of this third corona, however, needs further examination, 
because the soft excess may alternatively be modeled in terms of 
relativistically-smeared ionized reflection (e.g., Fabian et al. 2004) 
or absorption by disk winds (e.g., Cierlinski \& Done 2004). 

%%%%%%%%%%%table 1%%%%%%%%%%%%
\begin{deluxetable*}{ccccccc}[B]
\tablewidth{0pt}
%\rotate
\tablecaption{X-ray observations by \textit{Suzaku}}
\tablehead{
\colhead{Epoch} & \colhead{Observation Start} 
                & \colhead{End} 
                & \colhead{Middle}
                & \colhead{Exposure Time}\\
                & \colhead{(UT)} & \colhead{(UT)} & \colhead{(MJD)}
                & \colhead{(ksec)} }
\startdata
  1 & 2013 Apr  9 23:13:20 &  Apr 11 01:06:16 & 56392.51 & 51\\
  2 & 2013 Apr 27 00:17:13 &  Apr 27 10:42:22 & 56409.27 & 19\\
  3 & 2013 May 12 00:22:24 &  May 13 02:30:23 & 56424.58 & 50\\
  4 & 2013 May 23 03:32:08 &  May 24 07:05:07 & 56435.72 & 51\\
  5 & 2013 May 29 11:02:50 &  May 30 15:15:14 & 56442.05 & 54\\
  6 & 2013 Nov  4 06:15:13 &  Nov  5 05:10:17 & 56600.74 & 46\\
  7\tablenotemark{a} & 2014 Apr 7 16:54:26 &  Apr 8 12:00:24 & 56755.12 & 52
\enddata
\label{tab:Xobs}
\vspace{0.2cm}
\tablenotetext{a}{Only the XIS data are utilized.}
\end{deluxetable*}
%%%%%%%%%%%table 1%%%%%%%%%%%%

%%%%%%%%%%%table 2%%%%%%%%%%%%
\begin{deluxetable*}{lcccccc}[b]
\tablewidth{0pt}
%\rotate
\tablecaption{Telescopes and instruments for the optical observations}
\tablehead{
\colhead{Telescope} & \colhead{Mirror Diameter} &
  \colhead{Instrument} & \colhead{Field of View} & \colhead{Pixel Scale} &
  \colhead{Observing Band} & \colhead{$n_{\rm obs}$\tablenotemark{a}} \\
                    & \colhead{(m)} &
                       & \colhead{(arcmin)} & \colhead{(arcsec pixel$^{-1}$)} &
                       & }
\startdata
  Pirka & 1.6  & MSI\tablenotemark{b} & $3.3\times 3.3$ & $0.39$ & $B$, $V$ & 86 \\
  MITSuME & 0.5  & MITSuME\tablenotemark{c} & $28\times 28$ & $1.64$ & $g'$, $R_{\rm C}$, $I_{\rm C}$  & 6   \\
  Kiso Schmidt & 1.5  & KWFC\tablenotemark{d} & $60\times 30$\tablenotemark{e} & $0.95$ & $B$, $V$ &32\\
  Nayuta &  2.0 & MINT\tablenotemark{f} &  $11\times 11$ & $0.32$ & $B$, $V$ & 31 \\
  Kanata & 1.5 & HOWPol\tablenotemark{g} & $\phi 15$ & $0.29$ & $B$, $V$ &31   
\enddata
\label{tab:optobs}
\vspace{0.2cm}
\tablenotetext{a}{The total number of observing nights.}
\tablenotetext{b}{Multi-Spectral Imager (MSI; Watanabe et al. 2012).}
\tablenotetext{c}{Multicolor Imaging Telescopes for Survey and Monstrous Explosions (MITSuME; Kotani et al. 2005).}
\tablenotetext{d}{Kiso Wide Field Camera (KWFC; Sako et al. 2012).}
\tablenotetext{e}{Only one of the eight CCDs installed in KWFC was used.}
\tablenotetext{f}{Multiband Imager for Nayuta Telescope (MINT; Ozaki et al. 2005).}
\tablenotetext{g}{Hiroshima One-shot Wide-field Polarimeter (HOWPol; Kawabata et al. 2008).\\ \\ \\ \\}
\end{deluxetable*}

%%%%%%%%%%%table 2%%%%%%%%%%%%

To clarify  geometry of materials
around a SMBH, 
correlations and time lags between emissions in different energy bands are useful. 
Recently, studies of reverberation between different X-ray bands revealed 
that soft X-rays and Fe-K$\alpha$ emission lines have positive time lags, by hundreds seconds,  
against hard X-rays, with an implication that some reprocessing materials are present  
near the hard X-ray emitters (e.g., Uttley et al. 2014).  
Likewise, 
X-ray and optical/ultraviolet (UV) correlations are useful 
to investigate coronal geometries around an accretion disk.
Because seed photons for the inverse Comptonization process 
are presumably provided via the optical/UV disk black body, 
variations in these low-energy bands can precede that of X-rays 
with a lag of $\sim$days (e.g., Nandra et al. 2000). 
When fluctuations of the mass accretion rate propagate 
inward from the accretion disk to the corona, the optical emission 
is also expected to precede X-rays, but on a viscous or thermalizing time 
scale which may be much longer than days (e.g., Uttley \& Casella 2014). 
On the other hand, the disk should be illuminated by the X-rays, 
so that X-ray variations can cause optical flux changes, 
producing delays of $\sim$days in the opposite sense (e.g., Krolik et al. 1991). 
In order to distinguish these cases, 
and to better understand the geometry of the coronae and accretion disk, 
we need to measure the sign and length of the optical vs. X-ray time lag, 
and quantify the strength of their correlations.

So far, a large amount of effort has been invested on coordinated X-ray
and optical/UV observations of a number of AGNs. 
However, we are far from achieving a unified view. 
For example, Seyfert galaxies including 
NGC 5548 (Uttley et al. 2003; Suganuma et al. 2006; McHardy et al. 2014), 
NGC 3783 (Arevalo et al. 2009), and Mrk 79 (Breedt et al. 2000) exhibited 
relatively strong correlations with optical lags by several days, while those 
including Ark 564 (Gaskell 2006) and NGC 3516 (Maoz et al. 2002) showed 
much poorer correlations. 
One possible cause of this variety, we speculate, is that the different X-ray primary 
continua have different correlations to the optical/UV signals, and the previous 
coordinated observations mixed up the effects from these multiple X-ray components. 
To overcome this difficulty, simultaneous observations of AGNs should be conducted, 
under a condition wherein the different X-ray components can be clearly identified and separated.

To derive the geometrical information from the multi-wavelength correlations, 
we performed an X-ray and optical simultaneous monitoring  during 2013--2014, 
by utilizing \textit{Suzaku} (Mitsuda et al. 2007) and 5 Japanese ground-based telescopes.  
The target of this campaign is the bright type 1.5 Seyfert galaxy NGC 3516. 
It has a redshift of $z=0.00884$ 
which corresponds to a distance of $D=41.3$ Mpc $=1.3\times10^{26}$ cm. 
Its SMBH is estimated to have a mass of
$M_{\rm BH} = 3.2 \times 10^7~M_{\odot}$ (Denney et al. 2010),
which yields
the Schwarzschild radius of $R_{\rm S} = 0.95\times 10^{13}$ cm.
The column density of the Galactic interstellar absorption toward NGC 3516 
is $N_{\rm H} = 4.08 \times 10^{20}$ cm$^{-2}$ (Dickey \& Lockman 1990),
and the optical Galactic extinction is $A_{B}=0.151$ mag
(NASA/IPAC Extragalactic Database based on Schlafly \& Finkbeiner 2011).
Although NGC 3516 was previously reported, by Maoz et al. (2002), 
to exhibit a poor X-ray and optical correlation, 
the result will change if we properly decompose its X-ray continuum 
using the technique of  Noda et al. (2011b, 2013b), 
because NGC 3516 is one of the prototypical objects 
to which this method has been successfully applied (Noda et al. 2013). 
Unless otherwise stated, errors shown in tables and figures in the present paper 
refer 1$\sigma$ errors.

%=================
\section{Observation and Data reduction}
%=================

%------------------------------------------
\subsection{X-ray}
%------------------------------------------

The X-ray monitoring observations of NGC 3516 were performed 
in the \textit{Suzaku} AO-8 cycle during 2013--2014. 
Specifically, NGC 3516 was observed with \textit{Suzaku} 
five times from 2013 April 10 to May 29, with intervals of $\sim 1$--2 weeks. 
Another pointing was carried out on November 7, and the last one on 2014 April 4;  
thus, total 7 observations, with various intervals from days to months. 
In the present paper, we call the 1st, 2nd, ... and the 7th observations 
epoch 1, epoch 2, ... and epoch 7, respectively. 
The exposure in each epoch was $\sim 50$ ksec, 
except for epoch 2 which has an exposure of $\sim 20$ ksec. 
In all these epochs, the XIS (Koyama et al. 2007) and HXD (Takahashi et al. 2007) 
were operated in their normal modes, 
and data of the XIS and HXD-PIN are utilized in the present paper, 
except for epoch 7, in which the HXD-PIN  count rate was too low to 
significantly detect the signals. 

The data of the XIS and HXD-PIN were processed via the software version 2.4. 
In the XIS analysis, we added the data of XIS0 and XIS3, and 
utilized it as FI data, while we did not use XIS1 due to its higher background. 
The XIS source events were extracted from a $120''$-radius circle with its center on the source. 
Background events were accumulated from a surrounding annular region 
with the inner radius of $180''$ and  the outer radius of $270''$.
The response and ancillary-response files were prepared via 
softwares in HEASOFT 6.14 called \textit{xisrmfgen} and \textit{xissimarfgen} 
(Ishisaki et al. 2007), respectively. 
The HXD data were prepared by the same way as that of the XIS. 
In the HXD analysis, Non X-ray Background (NXB) and Cosmic X-ray Background (CXB)
were estimated by standard models described by Fukazawa et al. (2009) 
and Gruber et al. (1999), respectively. 
They were subtracted from the on-source data.

%--------------------------
\subsection{Optical}
%--------------------------

The optical photometric monitoring observations
of NGC 3516 were performed by using 
the charge-coupled device (CCD) cameras 
installed at the Pirka, MITSuME, Kiso Shcmidt, Nayuta, 
and Kanata telescopes, of which major parameters 
are summarized in Table \ref{tab:optobs}.  
Simultaneous monitoring observations were performed
at the same 7 epochs as the X-ray observations by {\it Suzaku}.
In addition, from 2013 January to 2014 April, 
these telescopes densely monitored the source 
even without the simultaneous X-ray coverage, 
with a typical intervals of $\sim1$~day.
In the present paper, we present the $B$- and the $g'$-bands photometric data,
because the optical continuum emission of AGNs is generally
bluer than that of the host galaxy,
thus ensuring better sensitivity to the AGN signals. 
The images were reduced using IRAF\footnote{
IRAF is distributed by the National Optical Astronomy Observatories,
which are operated by the Association of Universities for Research in
Astronomy, Inc., under cooperative agreement with the National Science
Foundation.
} following the standard procedures of image reduction
for CCD detectors.

%==================================================
\section{Data Analysis and Results}
%==================================================

%----------------------------------------------------------------------------------------
\subsection{X-ray Light Curves and Spectra}
%-----------------------------------------------------------------------------------------

%%%%%%%%%%%%%%%%figure1%%%%%%%%%%%%%%%%%%%%%%%%
\begin{figure*}[t]
\epsscale{1}
\plotone{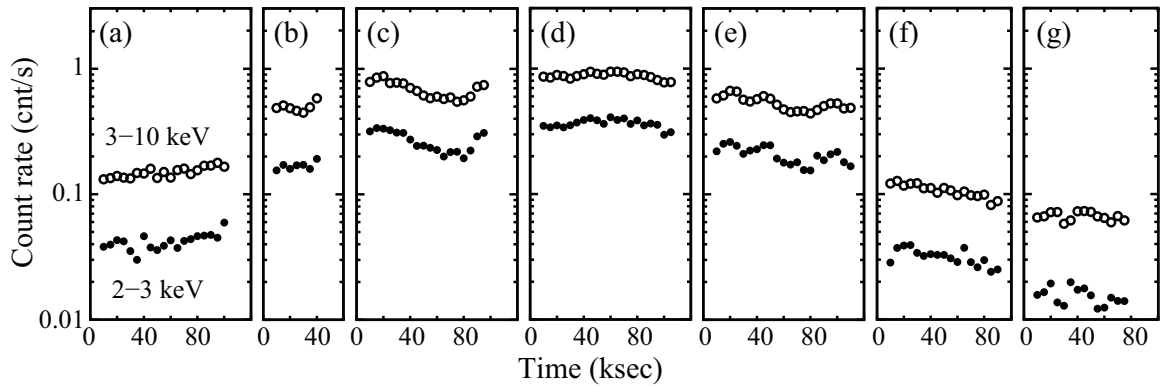}
\caption{XIS light curves of NGC 3516 in 
the 2--3 keV (open circles) and 3--10 keV (filled circles) bands, 
binned into 5 ksec, in epoch 1 (panel a), 2 (panel b), 3 (panel c), 4 (panel d), 5 (panel e), 
6 (panel f), and 7 (panel g). 
The error bars are all within $0.01$ cnt s$^{-1}$ in both bands, and are hence omitted. \\\\}
\label{fig:XIS_lc}
\end{figure*}
%%%%%%%%%%%%%%%%figure1%%%%%%%%%%%%%%%%%%%%%%%%

%%%%%%%%%%%%%%%%figure2%%%%%%%%%%%%%%%%%%%%%%%%

%%%%%%%%%%%%%%%%figure2%%%%%%%%%%%%%%%%%%%%%%%%
\begin{figure}[t]
\epsscale{1}
\plotone{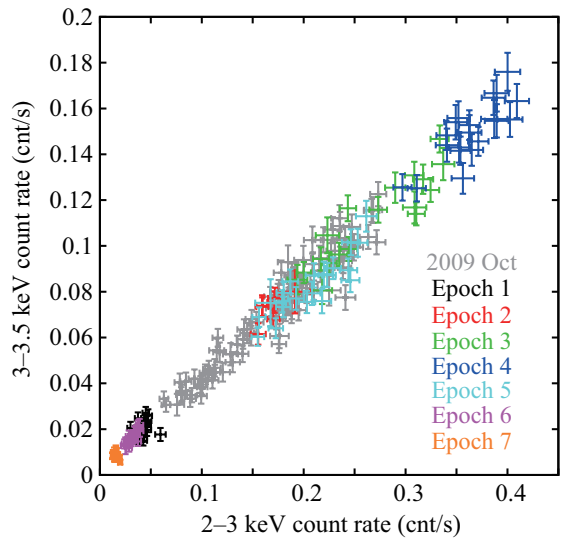}
\caption{A count-count plot between the 2--3 keV and 
 3--3.5 keV count rates, binned into 5 ksec. Black, red, green, blue, cyan, 
purple and orange indicate epoch 1, 2, 3, 4, 5, 6, and 7, respectively. 
Grey shows the data on 2009 October 28, which is same as those in Noda et al. (2013c), 
but corrected for the difference of the pointing position and some response 
changes from 2009 to 2013. \\\\}
\label{fig:Xccp}
\end{figure}
%%%%%%%%%%%%%%%%figure2%%%%%%%%%%%%%%%%%%%%%%%%

%%%%%%%%%%%%%%%%figure3%%%%%%%%%%%%%%%%%%%%%%%%

%%%%%%%%%%%%%%%%figure3%%%%%%%%%%%%%%%%%%%%%%%%
\begin{figure*}[t]
\epsscale{1}
\plotone{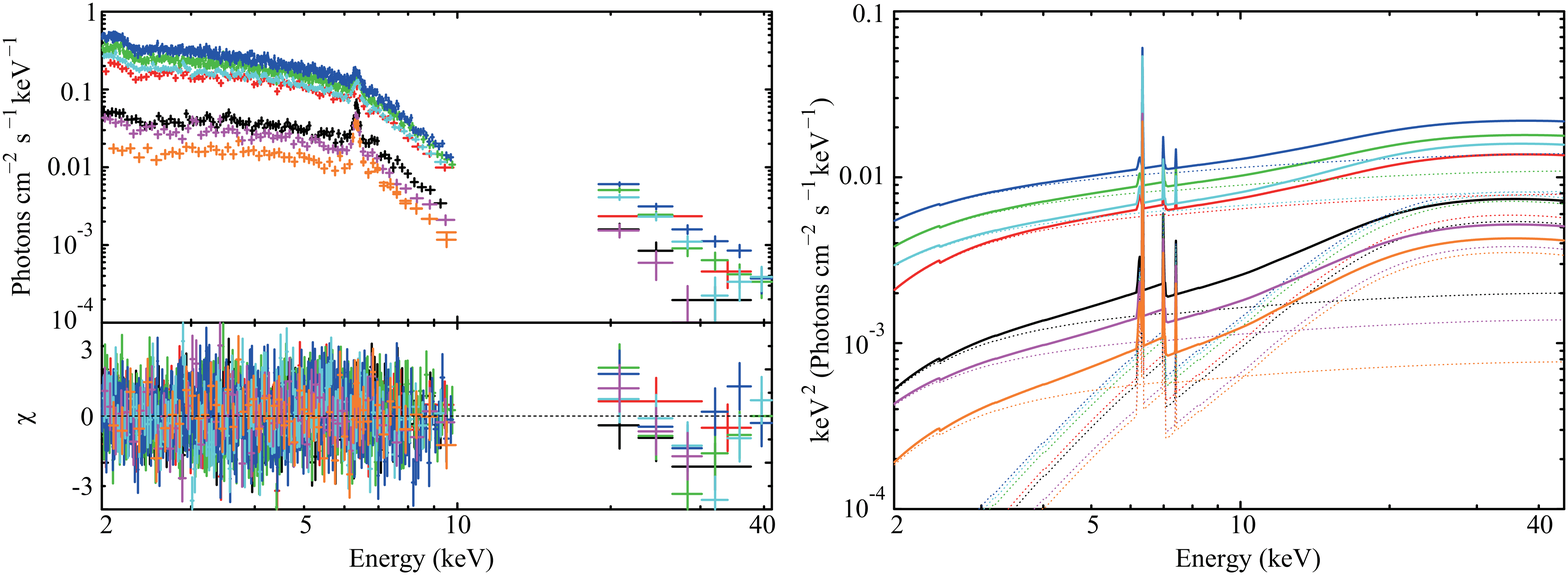}
\caption{The 2--45 keV time-averaged spectra, of which the epochs are specified 
by the same color as in Figure \ref{fig:Xccp}. They are fitted simultaneously with the model 
of \texttt{wabs * (cutoffpl + pexmon)} in XSPEC12. 
Left panel shows the spectra including the XIS and HXD-PIN detector responses (top) 
and residuals from the model (bottom). Right panel shows unfolded best-fit model spectra 
in a $\nu F_{\nu}$ form.  }
\label{fig:Xspectra}
\end{figure*}
%%%%%%%%%%%%%%%%figure3%%%%%%%%%%%%%%%%%%%%%%%%

Figure \ref{fig:XIS_lc} shows 
light curves in the  2--3 keV and 3--10 keV bands in all the 7 epochs. 
While we can see gradual flux changes, 
the source did not show short-term variations, on time scales of  $\sim$hours,  
that were observed by \textit{XMM-Newton} in 2006 (Mehdipour et al. 2010). 
The 2--3 keV light curves are clearly synchronized with that in the 3--10 keV band.  
The highest flux was obtained in epoch 4, while the lowest in epoch 7, 
with a peak to peak variation amplitude reaching a factor of $\sim 20$.  
To quantify the flux variability, we made, in Figure \ref{fig:Xccp}, 
a count-count plot between count rates in the 2--3 keV and 3--3.5 keV bands, 
where these objects generally exhibit rather high variability (Noda et al. 2013a).  
The count-count plot of NGC 3516 derived in 2009 October (Noda et al. 2013c) 
is also shown for reference, after correcting the data for a pointing position difference 
and long-term changes in the detector response. 
Surprisingly, all the data points in 2013--2014 
line up with those in 2009, thus defining 
an almost linear correlation between the two bands. 
This means that the variable component 
in those energy bands observed in 2013--2014 has 
nearly the same spectral shape to that detected in 2009.

In energy bands higher than 3.5 keV, a reflection component with 
a prominent Fe-K$\alpha$ emission line at $\sim 6.4$ keV is 
expected  to become more dominant than in energies below 3.5 keV.  
Therefore, this variable component should be separated from the 
reflection via spectral fitting. 
Although the reflection component is considered almost stationary 
within a week (Noda et al. 2013b), it can vary on timescales 
of months, because a broad line region and/or a dusty torus, 
where the reflection emission possibly originates, are known to locate at 
several tens--hundreds light days (e.g., Koshida et al. 2014). 
Accordingly, we treat the reflection signal as a variable component. 

%%%%%%%%%%%table 3%%%%%%%%%%

\begin{deluxetable*}{cccccccccc}[t]
\tablewidth{0pt}
\tabletypesize{\footnotesize}
%\tabletypesize{\small}
%\rotate
\tablecaption{Parameters obtained by the simultaneous fitting to 
 all the 2--45 keV time-averaged spectra}
\tablehead{
  \colhead{Component} & \colhead{Parameter}  & 
  \colhead{Epoch 1} & \colhead{Epoch 2} & \colhead{Epoch 3} &
  \colhead{Epoch 4} & \colhead{Epoch 5} & \colhead{Epoch 6} &
  \colhead{Epoch 7}}
\startdata
\texttt{wabs} &$ N_{\rm H}$\tablenotemark{a}
   			&$1.93^{+0.14}_{-0.15}$
			&$1.86^{+0.10}_{-0.11}$ 
			&$1.21^{+0.06}_{-0.07}$
			&$0.94^{+0.05}_{-0.06}$
                          &$1.16 \pm 0.07$
                          &$1.52^{+0.16}_{-0.17}$
                          &$2.10^{+0.21}_{-0.25}$\\[1.5ex]
   
 \texttt{cutoffpl} &$\Gamma_{\rm PL}$%
   			&\multicolumn{7}{c}{$1.75^{+0.01}_{-0.02}$}\\
                      
  &  $E_{\rm cut,~PL}$~(keV)
                      &\multicolumn{7}{c}{$150$ (fix)}\\

     &$N_\mathrm{PL}$\tablenotemark{b}%
                     &$1.00^{+0.02}_{-0.03}$
                     &$3.94^{+0.11}_{-0.13}$
                     &$5.44^{+0.09}_{-0.12}$
                      &$6.93^{+0.12}_{-0.15}$
                      &$4.10^{+0.07}_{-0.09}$
                      &$0.69^{+0.02}_{-0.03}$
                      &$0.39\pm0.02$\\				
 
    & $F_{\rm PL}$\tablenotemark{c}
     		& $0.37 \pm 0.01$
		& $1.44^{+0.04}_{-0.05}$
		& $1.99^{+0.03}_{-0.04}$
		& $2.53^{+0.04}_{-0.05}$
		& $1.50^{+0.02}_{-0.03}$
		& $0.25 \pm 0.01$
		& $0.14 \pm 0.01$ \\[1.5ex]

 \texttt{pexmon} &  $\Gamma_\mathrm{ref}$%
    			&\multicolumn{7}{c}{=$\Gamma_{\rm PL}$}\\
                          
  	&$E_{\rm cut,~ref}$~(keV)
                      &\multicolumn{7}{c}{= $E_{\rm cut,~PL}$}\\

 & $f_\mathrm{ref}$%
    			&\multicolumn{7}{c}{$1$ (fix)}\\
                          
 & $A_{\rm Fe}$~($Z_{\odot}$)
                          &\multicolumn{7}{c}{$1$ (fix)}\\
                          
 & $N_\mathrm{ref}$\tablenotemark{d}%
    			&$3.88^{+0.25}_{-0.26}$
			&$4.24^{+0.51}_{-0.50}$
			&$5.17^{+0.42}_{-0.44}$
                          &$5.88^{+0.47}_{-0.49}$
                          &$5.60^{+0.40}_{-0.41}$
                          &$2.73 \pm 0.21$
                          &$2.51 \pm 0.18 $\\
                          
& $F_\mathrm{ref}$\tablenotemark{e}%
    			&$1.10 \pm 0.07$
			&$1.20^{+0.15}_{-0.14}$
			&$1.46 \pm 0.12$
                          &$1.67^{+0.13}_{-0.14}$
                          &$1.59^{+0.11}_{-0.12}$
                          &$0.78 \pm 0.06$
                          &$0.71 \pm 0.05$\\[1.5ex]

& $F_\mathrm{total}$\tablenotemark{f}%
    			&$0.48 \pm 0.04$
			&$1.56 \pm 0.07$
			&$2.14^{+0.06}_{-0.07}$
                          &$2.70 \pm 0.07$
                          &$1.66^{+0.05}_{-0.06}$
                          &$0.33 \pm 0.03$
                          &$0.21\pm 0.02$\\[1.5ex]     
     
\hline

  & $\chi^{2}$/d.o.f.  &\multicolumn{7}{c}{1184.0/1113}
\enddata
\label{tab:specfit}
\tablecomments{The errors refer to $1\sigma$ confidence ranges.}
\tablenotetext{a}{Equivalent hydrogen column density in  $10^{22}$ cm$^{-2}$.}
\tablenotetext{b}{The power-law normalization at 1 keV, in units of $10^{-3}$~photons~keV$^{-1}$~cm$^{-2}$~s$^{-1}$~at 1 keV.}
\tablenotetext{c}{The 2--10 keV flux of the PL component without being absorbed, in units of $10^{-11}$~erg~cm$^{-2}$~s$^{-1}$.}
\tablenotetext{d}{The \texttt{pexmon} normalization at 1 keV, in units of $10^{-3}$~photons~keV$^{-1}$~cm$^{-2}$~s$^{-1}$~at 1 keV.}
\tablenotetext{e}{The 2--10 keV flux of the reflection component without being absorbed, in units of $10^{-12}$~erg~cm$^{-2}$~s$^{-1}$.}
\tablenotetext{f}{The 2--10 keV total flux without being absorbed in units of $10^{-11}$~erg~cm$^{-2}$~s$^{-1}$, calculated as a sum of  $F_\mathrm{PL}$ and  $F_\mathrm{ref}$. \\ }
\end{deluxetable*}
%%%%%%%%%%%table 3%%%%%%%%%%%%

As presented in Figure \ref{fig:Xspectra}, we produced seven  
time-averaged and background-subtracted spectra, 
one from each epoch,  and performed a simultaneous model fitting to them. 
Because of the presence of diffuse X-ray emissions in the host galaxy 
(e.g., Constantini et al. 2000; George et al. 2002), 
and effects of absorption variations (e.g., Turner et al. 2011), 
we ignored energy bands lower than 2 keV, and utilized the 2--45 keV energies in each spectrum. 
The employed spectral model is \texttt{wabs * (cutoffpl + pexmon)} in XSPEC12, 
which is consistent with that utilized in Noda et al. (2013a, c). 
Here, \texttt{wabs} is a model for the photoelectric absorption 
(Morrison and McCammon 1983), 
and its column density $N_{\rm H}$ was allowed to differ among the epochs.
The \texttt{cutoffpl} model represents the PL continuum with 
a high energy exponential rolloff, utilized to emulate the inverse Comptonization radiation. 
The reflection continuum and the Fe-K$\alpha$ emission line, both produced by this continuum, 
are reproduced by \texttt{pexmon} (Nandra et al. 2007). 

In the fitting, the photon index $\Gamma$ 
was tied among the epochs (because of Figure 2), and left free.  
The cutoff energy $E_{\rm cut}$ was fixed at 150 keV based on  
the typical value reported by Malizia et al. (2014), 
while the normalization $N_{\rm PL}$ was left free in each epoch.
For \texttt{pexmon}, 
$\Gamma$ and $E_{\rm cut}$ of the incident PL were 
tied to those in \texttt{cutoffpl}, and  the Fe abundance $A_{\rm Fe}$, 
inclination angle $I$, and the reflection fraction $f_{\rm ref}$ were 
fixed at 1 Solar, 60$^{\circ}$, and 1, respectively. 
The normalization $N_{\rm ref}$ of the \texttt{pexmon} component 
was allowed to differ among the epochs. 
The $N_{\rm ref}$ value is determined almost solely by the Fe-K$\alpha$ line intensity,  
because the ratio between the Fe-K$\alpha$ line intensity and the reflection continuum 
in \texttt{pexmon} does not change when $E_{\rm cut}$, $A_{\rm Fe}$, and $I$ are all fixed. 
As summarized in Figure \ref{fig:Xspectra} and Table \ref{tab:specfit}, 
the simultaneous fitting was successful with $\chi^2$/d.o.f.=1184.0/1113.
As expected from Figure \ref{fig:Xccp}, all the spectra have been reproduced 
by a hard PL continuum having $\Gamma \sim 1.75$, which is consistent with that in 2009 
($\Gamma = 1.72^{+0.08}_{-0.12}$),   
together with a reflection component accompanied by a prominent Fe-K$\alpha$ emission line. 
Although $N_{\rm H}$ slightly varied, 
the low-energy shapes of the time-averaged spectra 
were not so strongly affected (see Figure \ref{fig:Xspectra}).

%%%%%%%%%%%%%%%%%figure4%%%%%%%%%%%%%%%%%%%%%%%
\begin{figure*}[t]
\epsscale{0.9}
\plotone{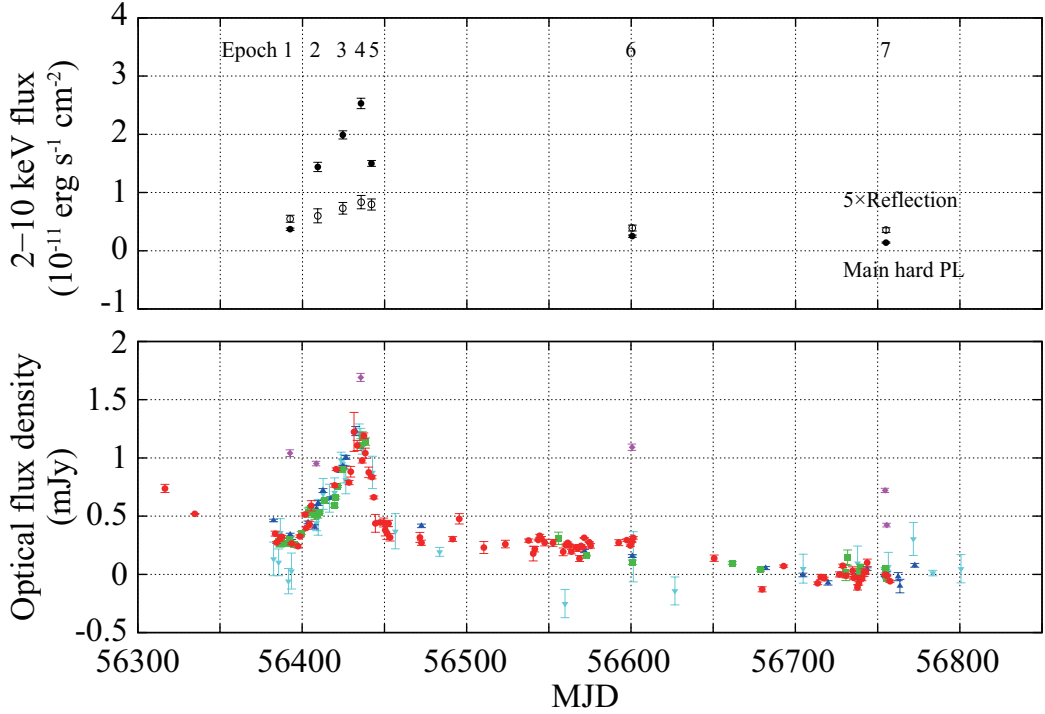}
\caption{(Top) The 2--10 keV flux light curve of the main hard PL continuum (filled circles),   
and that of the reflection component including the Fe-K$\alpha$ line, 
after multiplying by 5 (open circles), both corrected for absorption. 
(Bottom) A light curve of the $B$-band flux density of the NGC 3516 nucleus, 
derived by applying the differential image photometry to the data of 
the Pirka (red), Kiso Schmidt (green), Nayuta (blue), and the Kanata telescope (cyan). 
The $g'$-band flux density obtained by the MITSuME telescope (magenta) is also  
plotted, with an offset of 0.5 mJy added after being scaled (see text). \\ }
\label{fig:XOLC}
\end{figure*}
%%%%%%%%%%%%%%%%%figure4%%%%%%%%%%%%%%%%%%%%%%

%%%%%%%%%%%%%%%%figure5%%%%%%%%%%%%%%%%%%%%%%%%
\begin{figure*}[t]
\epsscale{0.9}
\plotone{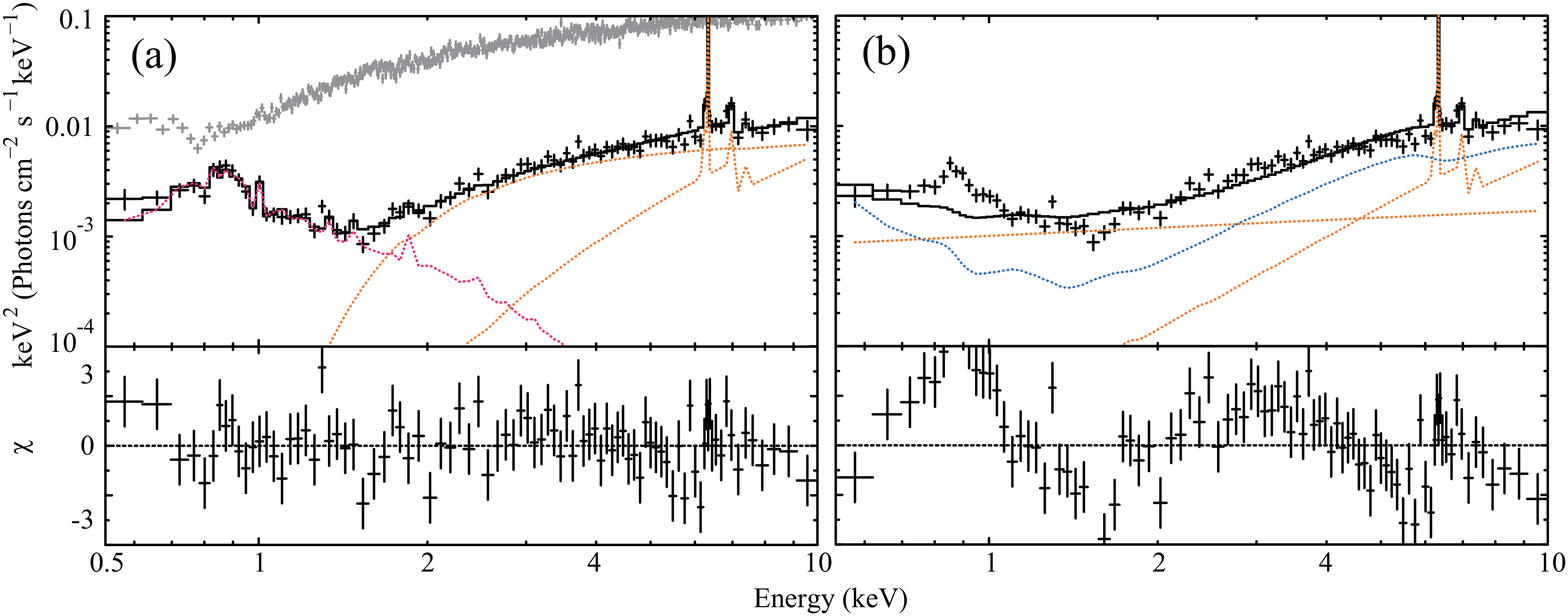}
\caption{Examples fo X-ray spectral analysis including the soft energy band below 2 keV. 
Black shows the 0.5--10 keV time-averaged spectrum at epoch 7 in a $\nu F_{\nu}$ form, 
fitted by \texttt{wabs * (cutoffpl + pexmon) + apec} (panel a) and  
\texttt{wabs * (cutoffpl + pexmon + kdblur * reflionx)} (panel b). 
Grey in panel (a) shows that at epoch 4 without fitting.  
In both panels, orange spectral components are same as those in 
Figure \ref{fig:Xspectra} except for absorption strength, 
while red in panel (a) and blue in panel (b) 
represent a galactic thin-thermal emission modeled by \texttt{apec} 
and a relativistically-blurred ionized reflection component modeled by \texttt{kdblur * reflionx}, 
respectively.  \\ \\
}
\label{fig:LowEneSpec}
\end{figure*}
%%%%%%%%%%%%%%%%figure5%%%%%%%%%%%%%%%%%%%%%%%%

The highest and lowest 2--10 keV fluxes were 
$\sim2.7\times10^{-11}$~erg s$^{-1}$ cm$^{-2}$ in epoch 4 
and $\sim 0.3 \times10^{-11}$~erg s$^{-1}$ cm$^{-2}$  in epoch 7, respectively.  
They are in a range of just 7\%--70\% of the  2--10 keV flux 
of $\sim 4.0 \times 10^{-11}$~erg s$^{-1}$ cm$^{-2}$  averaged over 1997--2002  
(\textit{RXTE} AGN Timing \& Spectral Database; Maoz et al. 2002);  
thus NGC 3516 was in an X-ray faint phase during the present observations. 
Table \ref{tab:specfit} further gives the 2--10 keV fluxes of the hard PL continuum 
and the reflection components, calculated separately, after removing the absorption. 
The 2--10 keV light curves of the two spectral components, derived in this way, 
are presented in Figure \ref{fig:XOLC}(top). 
The time of each data point refers to the middle epoch (in MJD) of that observation 
as given in Table \ref{tab:Xobs}. 
Interestingly, 
the flux of the reflection component significantly changed by a factor of 2 
on a time scale of several months. 
However, the hard PL component varied almost by an order of magnitude or more in amplitude. 
Clearly, the large intensity change in Figure \ref{fig:Xccp} was mostly carried by the hard PL variation. 
The largest flux change of the hard PL emission in 2--10 keV  is 
$\sim 2.4 \times10^{-11}$ erg s$^{-1}$ cm$^{-2}$ in difference, 
and a factor of $\sim 18$ between epoch 4 and 7.

For further information about the AGN activity in NGC 3516, 
we also performed the spectral analysis including the energy band below 2 keV.  
For this purpose we selected epoch 7, where the AGN was faintest. 
In fact, as shown in Figure \ref{fig:LowEneSpec}, 
the spectrum on this occasion exhibits a prominent soft X-ray excess at $\lesssim 2$ keV, 
which is much less prominent in the other epochs.  
In order to examine if the soft excess structure originates from the AGN activity, 
we fitted the 0.5--10 keV spectrum at epoch 7 by two contrasting models; 
(a) \texttt{wabs * (cutoffpl + pexmon) + apec} and (b)  
\texttt{wabs * (cutoffpl + pexmon + kdblur * reflionx)}. 
In model (a), the soft excess is explained by \texttt{apec} (Smith et al. 2001)
which represents thin-thermal plasma emission from the host galaxy,  
while in model (b) by a \texttt{kdblur * reflionx} model (Laor 1991; Ross \& Fabian 2005)
which describes relativistically-smeared reflection continuum. 
As a result, the fit with model (a) became successful with 
$\chi^2/\textrm{d.o.f.}=100.3/84$ (Figure \ref{fig:LowEneSpec}a), 
in which  the soft excess, apparently involving emission lines, 
is reproduced by the plasma emission model. 
On the other hand, the fits with model (b) was unsuccessful  
with $\chi^2/\textrm{d.o.f.}=355.9/82$, 
mainly due to lack of the emission lines at $\sim 0.85$ keV 
and a convex data shape in the 2--5 keV band (Figure \ref{fig:LowEneSpec}b). 
Therefore, the spectrum below 2 keV is
dominated by the host galaxy emission, at least in epoch 7.
This justifies us to limit the spectral studies to the energies above $\sim2$ keV. 
Of course, the AGN emission, when bright, probably dominates down to 
$\sim 0.5$ keV, but is strongly affected by ionized-absorption features at $\sim 0.8$ keV, 
as shown in Figure \ref{fig:LowEneSpec}(a).

%%%%%%%%%%%%%%%%%figure5%%%%%%%%%%%%%%%%%%%%%%%
\begin{figure}[t]
\epsscale{0.9}
\plotone{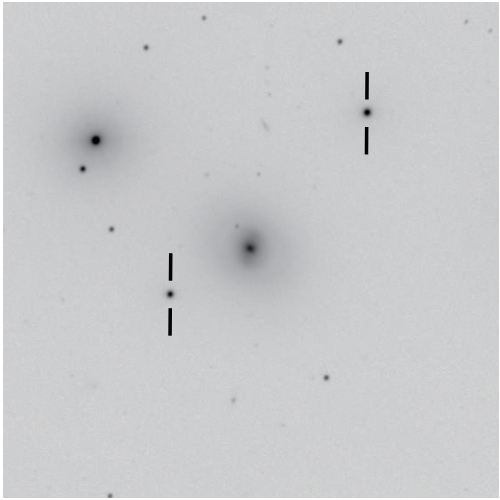}
\caption{An example of $B$-band reference image of NGC 3516
for the DIP analysis of the MINT/Nayuta telescope data.
It was obtained by stacking 65 images with an exposure of 60 seconds each
taken on 2014 April 7, when the FWHM of the PSF was 1.8 arcsec.
The two reference stars, marked by bars, 
were used for matching the photometric intensity and the PSF,
and for the relative photometry of the AGN flux.
North is up, east is left, the field of view of
the image displayed is $6\times 6$ arcmin$^2$, and
the image intensity levels are displayed in a logarithmic scale.
The NGC 3516 nucleus is near the image center. \\
}
\label{fig:NGC3516_B_MINT_6x6arcmin_log.ps}
\end{figure}
%%%%%%%%%%%%%%%%%figure5%%%%%%%%%%%%%%%%%%%%%%

%%%%%%%%%%%%%%%%%figure6%%%%%%%%%%%%%%%%%%%%%%%
\begin{figure}[t]
\epsscale{1}
\plotone{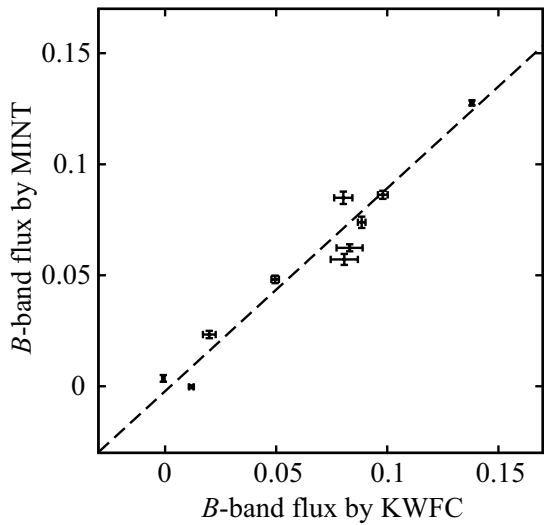}
\caption{
The $B$-band fluxes obtained by the KWFC/Kiso Schmidt telescope, 
compared with those from the MINT/Nayuta telescope on the same 
observing nights. 
The unit of flux is that of the average magnitude of 
the two nearby reference stars.
The dashed line represents the best fit linear regression
with an error added to the photometric errors by root-sum-square
for the reduced $\chi^{2}$ to achieve unity.\\
}
\label{fig:KWFC_MINT_ff.ps}
\end{figure}
%%%%%%%%%%%%%%%%%figure6%%%%%%%%%%%%%%%%%%%%%%

%--------------------------------------------------------
\subsection{Differential Imaging Photometry and Optical Light Curves}
%-------------------------------------------------------

%%%%%%%%%%%%%%%%%%figure7%%%%%%%%%%%%%%%%%%%
\begin{figure*}[t]
\epsscale{0.95}
\plotone{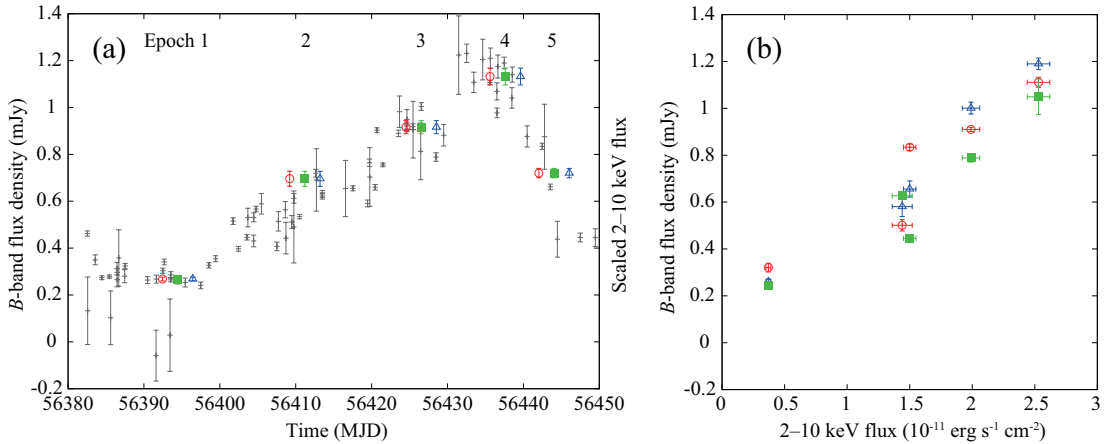}
\caption{(a) Zoomed light curves of the $B$ band (grey) and the 2--10 keV 
hard PL component (red circles) from epoch 1 to 5. 
Green squares and blue open triangles show the X-ray light curves 
which are purposely delayed by $+2$ and $+4$, respectively. 
The X-ray flux amplitude was scaled to match that in the $B$ band.
(b) A plot between the 2--10 keV hard PL flux
and the $B$-band flux density,  
wherein an artificial time delay is applied to the X-ray data by 
0 day (red circle), 2 days (green box), and 4 days (blue triangle). \\ }
\label{fig:XOLC1-5}
\end{figure*}
%%%%%%%%%%%%%%%%%%figure7%%%%%%%%%%%%%%%%%%%%

According to changes of the atmospheric seeing during an observation,
the flux of an AGN within an aperture changes differently
from that of the host galaxy,
leading to a significant uncertainty
in measuring the flux variation of a faint AGN hosted by a bright galaxy.
To minimize such uncertainty in the photometry of the NGC 3516 nucleus, 
we therefore performed Difference Image Photometry (DIP; Crotts 1992; Tomaney \& Crotts 1996).
As presented in Figure \ref{fig:NGC3516_B_MINT_6x6arcmin_log.ps} for example, 
a reference image was created by stacking many images
obtained at the same night with good seeing condition.
Then, we matched the position, the photometric intensity,
and the point-spread function (PSF) of the reference image
to those of each individual image and subtracted the former from the latter.
Two nearby field stars, located at
(Ra, Dec) $=$ (11:06:28.51, +72:35:46.2) and (11:07:00.41, +72:33:33.3),
were used as the reference to match the photometric intensity and the PSF.
The IRAF {\it psfmatch} task was used for matching the PSFs.
After that, for each individual image, 
the residual flux at the center of the galaxy was measured
relative to the two nearby reference stars, 
with a circular aperture of $\phi= 4\times $full-width at half maximum (FWHM)
of the PSF in diameter. 
Finally, the residual flux data at the same observing night were averaged
to obtain the flux difference of the NGC 3516 nucleus at that epoch
with respect to the reference image.
The photometric error was estimated from
the ensemble scatter of the DIP fluxes 
and the number of the images obtained at the same night.
These procedures were applied to the data set of each telescope, individually.

Because the observing epochs of the reference images used in the DIP analysis
depend on the telescopes,
there can be systematic offsets between them.
Systematic differences in the scaling factor among them
are also possible because of the differences in the filter color term.
To match the offset and the scaling factor among the four telescopes, 
we therefore performed a linear regression analysis to
their $B$-band flux data sets of the same observing nights.  
An example, between two telescopes, is presented in Figure \ref{fig:KWFC_MINT_ff.ps}. 
The reduced $\chi^2$ of the linear regression was much larger than unity
when only the photometric errors were incorporated. 
Because this suggests the presence of some systematic errors, 
we added an error $\sigma_{\rm add}$ to the photometric errors 
in quadrature, 
and regarded it as the systematic error of the photometry.
By requiring the reduced $\chi^2$ to become unity, 
$\sigma_{\rm add} \sim 0.06$ mJy was obtained.

The $B$- and $V$-band magnitudes of the two reference stars
were calibrated relative to those of the more distant field stars
whose magnitudes are presented in Sakata et al. (2010); 
these are based on the wide-field image data of the KWFC,  
in which both stars were observed simultaneously.
The $g'$-band magnitudes of the reference stars were estimated
from their $B$- and $V$-band magnitudes according to Jordi et al. (2006).

The light curves of the NGC 3516 nucleus, thus derived in the optical $B$ and $g'$ bands, 
are presented in the bottom panel of Figure \ref{fig:XOLC}, 
and the data are listed in Table \ref{tab:optical_fluxes}, 
without correction for the Galactic extinction. 
They are both the differential fluxes with respect to the reference image
obtained from the image data on 2014 April 08 (UTC), 
and the error bars of the data points do not include $\sigma_{\rm add}$. 
The total numbers of the photometric data points are 180 and 6 for
the $B$ and the $g'$ bands, respectively.
As shown in Figure \ref{fig:XOLC},
the $B$-band flux varied rather 
similarly to that of the hard PL component in the 2--10 keV band.
Although the number of the data points are small,
the $g'$-band flux variation also followed them.
The larger relative scatters of the $g'$-band data points 
are caused by the small telescope size and the large pixel scale of
the camera (MITSuME) by which the data were obtained.
These results indicate that 
the optical continuum and the 2--10 keV PL component 
varied in a good correlation with each other.
Between the peak and bottom, the $B$-band flux density varied by $\sim 1.2$ mJy.
After correcting for the Galactic extinction and the optical extinction of the NGC 3516 nucleus
($A_{B}\sim 1.3$ mag in total), 
it becomes $\sim 4.0$ mJy, corresponding to
$\sim 2.7\times 10^{-11}$ erg s$^{-1}$ cm$^{-2}$ in $\nu F_{\nu}$ units 
\footnote{
There are different estimates for the optical extinction of the NGC 3516 nucleus:
%based on the spectral shape of the UV-optical continuum emission and the Balmer
%decrement measured from the broad emission lines:
Cackett et al. (2007) estimated it at $E(B-V)=0.15$--$0.16$ mag,
which can be converted to $A_{B}=0.62$--$0.66$ mag,
and Denney et al. estimated it at $A_{B}=1.68$--$1.72$ mag.
We adopted their average here.
Although the $N_{\rm H}\sim 1-2\times 10^{22}$ cm$^{-2}$ derived from
the X-ray absorption suggests much larger optical extinction of $A_{B}\sim 10$ mag,
AGNs often show smaller optical extinction than
that converted from the X-ray $N_{\rm H}$
(e.g., Burtscher et al. 2015).
}.

%%%%%%%%%%%table 4%%%%%%%%%%
\begin{deluxetable}{lcccc}[t]
\tablewidth{0pt}
\tablecaption{Optical Fluxes}
\tablehead{
\colhead{Observation Date}     & \colhead{Observatory\tablenotemark{a}} &
\colhead{Filter}               & \colhead{Flux\tablenotemark{b}} & \colhead{Flux Error} \\
\colhead{(MJD)}                & \colhead{} &
                               & \colhead{(mJy)} & \colhead{(mJy)}}
\startdata
56316.544 & P & $B$  & 0.738 & 0.035 \\
56334.674 & P & $B$  & 0.519 & 0.003 \\
56382.554 & N & $B$  & 0.462 & 0.011 \\
56382.588 & H & $B$  & 0.133 & 0.144 \\
56383.576 & P & $B$  & 0.350 & 0.021 \\
56384.454 & P & $B$  & 0.274 & 0.008 \\
56385.448 & K & $B$  & 0.278 & 0.007 \\
56385.611 & H & $B$  & 0.103 & 0.115 \\
\nodata & \nodata & \nodata & \nodata  \\
56392.487 & K & $B$  & 0.303 & 0.011 \\
56392.625 & M & $g'$ & 0.541 & 0.029 \\
56392.686 & N & $B$  & 0.341 & 0.012 \\
\nodata & \nodata  & \nodata & \nodata  
\enddata
\label{tab:optical_fluxes}
\tablecomments{(This table is available in its entirety in a machine-readable
form in the online journal. A portion is shown here for
guidance regarding its form and content.)}
\tablenotetext{a}{Observatory code : 
P $=$ the MSI at the Pirka telescope, 
M $=$ the MITSuME, 
K $=$ the KWFC at the Kiso Schmidt telescope, 
N $=$ the MINT at the Nayuta telescope, and
H $=$ the HOWPol at the Kanata telescope.
}
\tablenotetext{b}{
The flux difference with respect to the reference image
on 2014 April 08 (UTC).
}
\end{deluxetable}

%%%%%%%%%%%table 4%%%%%%%%%%

To complete this DIP analysis, the AGN flux at the faintest phase
during our observation was estimated. We applied an aperture
photometry to the stacked $B$-band image obtained on 2014 April 7
by the MINT attached on the Nayuta telescope (Figure 5), 
because it achieved, on this day, the best seeing among the four. 
Sakata et al. (2010) estimated the host galaxy flux in $B$ band
as $8.38 \pm 0.18$ mJy, 
\footnote{
Since Sakata et al. (2010) assumed the Galactic extinction of
$A_B=0.183$ mag according to Schlegel, Finkbeiner \& Davis (1998),
we converted their host galaxy flux to that assuming
$A_B=0.151$ mag adopted in this paper.
}
within an aperture diameter of $\phi = 8.3$ arcsec with the sky
reference of a $\phi = 11.1$--$13.9$ arcsec annulus. By applying
the photometry with the same parameters, we obtained the $B$-band
flux density of $8.44$ mJy after the correction for the Galactic
extinction, yielding the AGN flux of $0.1 \pm 0.2$ mJy.
Considering the various systematic errors in the photometry,
the $B$-band AGN flux at the faintest phase during the observations
is estimated as about a few times $0.1$ mJy, with a similar
amount of flux error.

%-----------------------------------------------------------------
\subsection{Time-Series Analysis on the Flux Variations of the Hard PL X-ray and the Optical Continuum}
%----------------------------------------------------------------

As shown in Figure \ref{fig:XOLC}, 
the optical flux changed almost simultaneously with
the 2--10 keV hard PL flux. 
In order to examine possible time lags between them,
we first focus on their expanded light curves from epoch 1 to 5, 
as presented in Figure \ref{fig:XOLC1-5}(a). 
Clearly, the AGN became the brightest in both bands at epoch 4. 
Also plotted in the same figure is the X-ray light curve shifted 
by $+2$ days and $+4$ days. 
Apparently, the X-ray light curve with a few days shift
shows the closest agreement with the $B$-band light curve,
suggesting that the variation of the X-rays
preceded that of the $B$ band by a few days.

Figure \ref{fig:XOLC1-5}(b) presents the correlation
of the X-ray flux with $0$, $+2$, and $+4$ days temporal shifts, 
against the $B$-band flux at the delayed epochs. 
The latter was estimated by averaging 
the $B$-band fluxes in close observations within $\pm$1~day. 
The correlation between those fluxes appears to be strongest
when the X-ray light curve is shifted by $+2$ days. 
This reconfirms  that the X-ray variations 
preceded that of the $B$ band by a few days.
Below, let us quantify the suggested time lag with two methods.

%%%%%%%%%%%%%%%%%%figure8%%%%%%%%%%%%%%%%%%%
\begin{figure}[t]
\epsscale{1.1}
\plotone{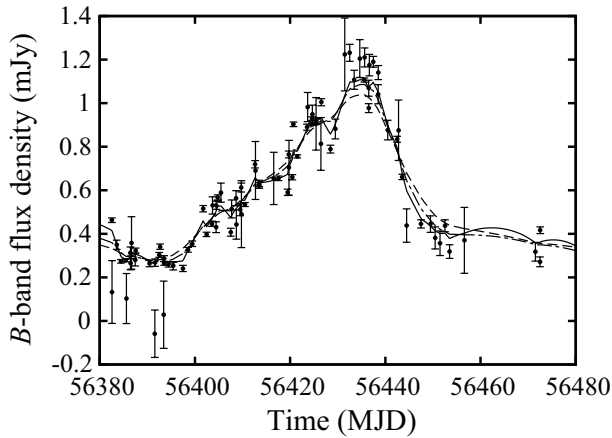}
\caption{
The observed $B$-band light curve (filled dots)
and its interpolations used for the time-series analysis.
The solid, dot-dashed, and dashed lines
represent 
the best fit light curves based on the DRW models
with $\sigma_{\rm add} = 0.00$ mJy, 0.03 mJy and 0.06 mJy, respectively.\\\\
}
\label{fig:BLC_sys}
\end{figure}
%%%%%%%%%%%%%%%%%%figure8%%%%%%%%%%%%%%%%%%%%

%%%%%%%%%%%%%%%%%%figure9%%%%%%%%%%%%%%%%%%%
\begin{figure*}[t]
\epsscale{1}
\plotone{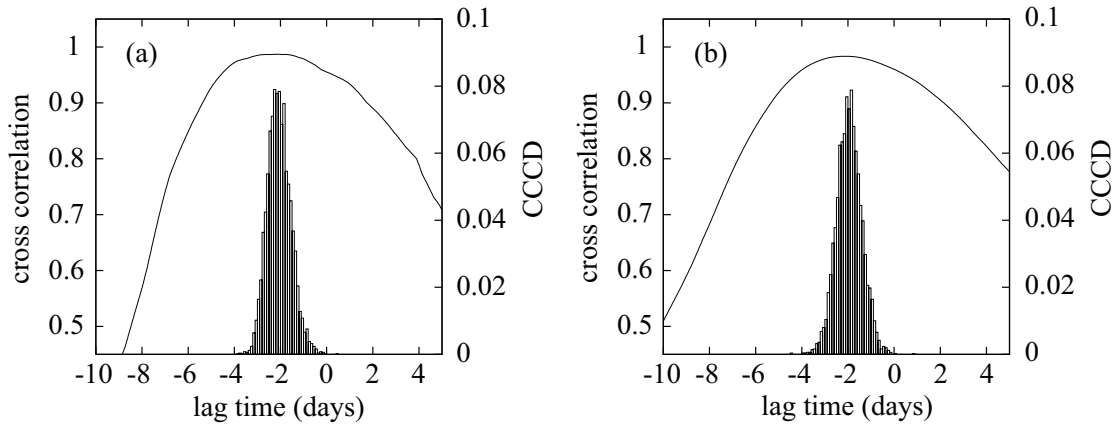}
\caption{
Results of the ICCF analysis between the flux variations
of the 2--10 keV hard PL and the $B$-band flux density.
The positive lag means that the optical leads X-rays. 
The solid line represents the cross correlation 
as a function of the time lag,
and the histogram shows the distribution of
the CCF centroid, $\tau _{\rm cent}$,
obtained by 5000 realizations of the FR/RSS simulation.
Panel (a) and (b) show the results with 
$\sigma_{\rm add} = 0.00$ mJy and 0.06 mJy, respectively.\\
}
\label{fig:CCCD}
\end{figure*}
%%%%%%%%%%%%%%%%%%figure9%%%%%%%%%%%%%%%%%%%

\subsubsection{Cross-correlation Analysis}

One method is the interpolated cross-correlation function (ICCF) method,
which has been widely used (White \& Peterson 1994; Peterson et al. 1998).
This is the same as the ordinary cross-correlation function (CCF), but either 
or both of the time-series data are interpolated, so that the CCF can be 
calculated even if the two data sets have 
different samplings, or when either or both are irregularly sampled. 

Since the monitoring cadence of the $B$-band data
was much higher than that of the X-ray data,
only the $B$-band light curve was interpolated,  
to make it nearly continuous. 
The interpolated $B$-band light curve was calculated
%%%
using a fitting code developed by Zu et al. (2011),
which assumes a damped random walk (DRW) model
for the flux variation.
%using the {\tt JAVELIN} software (formerly known as {\tt SPEAR}), 
%developed by Zu et al. (2011) assuming 
%a damped random walk (DRW) model for the flux variation.
%%%
The DRW model has been demonstrated to be a good statistical model
of flux variations of the UV-optical continuum emission of AGNs
(e.g., Kelly et al. 2009; Kozlowski et al. 2010;
MacLeod et al. 2010, 2012; Zu et al. 2013).
In this interpolation, the systematic error $\sigma_{\rm add}$ 
of the photometry, either as determined in \S 3.2 or somewhat changed, 
was added in quadrature.

%%%%%%%%%%%%%%%%%%figure10%%%%%%%%%%%%%%%%%%%
\begin{figure*}[t]
\epsscale{1}
\plotone{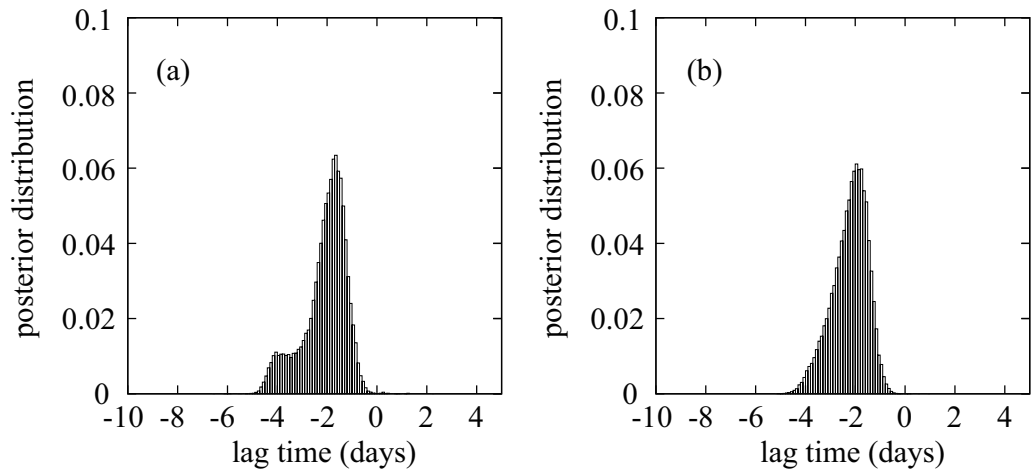}
\caption{
The same as Figure \ref{fig:CCCD}, but using the JAVELIN software. \\\\
}
\label{fig:javelin}
\end{figure*}
%%%%%%%%%%%%%%%%%%figure10%%%%%%%%%%%%%%%%%%%

Figure \ref{fig:BLC_sys} presents the observed $B$-band flux data and their interpolation.
Thus, the interpolated light curve does not follow the observed data 
when $\sigma_{\rm add} = 0.06$ mJy as determined in \S 3.2 is employed: 
the peak flux of the interpolated light curve was lower,
and its flux decrease after the peak was slower.
We generally found that a larger value of $\sigma_{\rm add}$ 
made the interpolated light curve smoother and less variable,
probably because
a larger $\sigma_{\rm add}$ would work as if applying a stronger low-pass filter. 
Therefore,
we performed the CCF analysis
using the $B$-band data with $\sigma_{\rm add}=0$ mJy in addition to 0.06 mJy 
to examine the possible uncertainty in lag caused by $\sigma_{\rm add}$.
As shown in Figure \ref{fig:BLC_sys},
the interpolated light curves then became
to follow the observed data much better
when $\sigma_{\rm add}$ is reduced.

The X-ray data of epoch 1--5 
and the optical data of MJD $=56316$--$56511$
were selected for the CCF analysis,
because the most remarkable flux variations were
present in those epochs,
and also because the $B$-band light curve data were sampled densely 
(Figure \ref{fig:XOLC} and \ref{fig:XOLC1-5}).
The calculated CCFs, in which the $B$-band data were smoothed using 
$\sigma_{\rm add} = 0$ mJy and 0.06 mJy, 
are presented in Figure \ref{fig:CCCD},
in which the time lag $\tau$ is defined to be positive if the optical precede X-rays. 
As shown in Figure \ref{fig:CCCD}, the CCF is peaked at about $\tau = -2$ days. 
This reconfirms the inference from Figure \ref{fig:XOLC1-5}, and implies that 
the optical variation is delayed from that in X-rays by $\sim 2$ days. 
The CCF value at the peak is very high, $>0.98$, 
% 0.0988 @0.02 mJy, 0.0985@0.03 mJy for epoch 1-5
as indicated by Figure \ref{fig:XOLC1-5}(b).

As quantitative measurement of $\tau$, 
we use the centroid of the CCF peak, $\tau_{\rm cent}$, 
which is computed from all neighboring points around the CCF peak
where CCF values are $> 0.95$ times that of the peak.
The uncertainty of $\tau_{\rm cent}$ is estimated 
using the model-independent Monte Carlo method of
flux randomization and random subset sampling (FR/RSS)
introduced by Peterson et al. (1998, 2004).
The FR method modifies the observed fluxes in each realization 
randomly within the errors assigned to the individual data points, 
and the RSS method randomly extracts the same number of data points
from the observed light curve allowing for multiple extraction. 
Then, the CCF and $\tau_{\rm cent}$ are calculated for each realization
in the same way, to produce the cross-correlation centroid distribution (CCCD).
The CCCDs calculated by 5000 realizations with the FR/RSS method
are presented in Figure \ref{fig:CCCD},
and the derived $\tau_{\rm cent}$ and its uncertainties are
listed in Table \ref{tab:lag}.
Thus, the results with $\sigma_{\rm add} = 0$ mJy and 0.06 mJy
agree well with each other. 
% 0.00 :   -2.085 +0.542 -0.476;  99% -0.729
% 0.06 :   -1.954 +0.548 -0.533;  99% -0.641
% average: -2.0195+0.5450-0.5045;  99% -0.72
On average, the lag has been estimated as
$\tau_{\rm cent}=-2.02^{+0.55}_{-0.50}$ days
 (1$\sigma $ error),
and the time lag is significantly non-zero, because the $99$\%-confidence limit is $\tau < -0.68$ days. 

% ICCF
%%%\bibitem[Gaskell \& Peterson (1987)]{1987ApJS...65....1G} Gaskell, C. M,
%%% \& Peterson, B. M., 1987, \apjs, 65, 1
% ICCF and CCPD
%\bibitem[White \& Peterson (1994)]{1994PASP..106..879W} White, R. J.,
% \& Peterson, B. M., 1994, \pasp, 106, 879
% ICCF and CCCD (see also Koratkar & Gaskell (1991), Penston 1991)
% FR/RSS
%\bibitem[Peterson et al. (1998)]{1998PASP..110..660P} Peterson, B. M.,
% Wanders, I., Horne, K., Collier, S., Alexander, T., Kaspi, S., \& Maoz, D.,
% 1998, \pasp, 110, 660
%\bibitem[Peterson et al. (2004)]{2004ApJ...613..682P} Peterson, B. M. et al.,
% 2004, \apj, 613, 682
%
%
% X-ray - optical multiband correlation -> Not Result, but Discussion 
%\bibitem[Edelson et al. (2015)]{} Edelson, R. et al.,
% 2015, \apj, 806, 129

\subsubsection{{\tt JAVELIN} Analysis}

The other method of the lag estimation is the {\tt JAVELIN} software developed by Zu et al. (2011),
which is widely employed not only in recent reverberation studies for
the optical broad emission lines and the thermal dust emission of AGNs
(e.g., Grier et al. 2012; Peterson et al. 2014; Koshida et al. 2014),
but also in the lag analysis between flux variations
of their X-ray emission and the UV-optical continuum emission
(Shappee et al. 2014; McHardy et al. 2014; Lira et al. 2015).
It explicitly builds a model of a response light curve
that is expressed as a convolution of a source light curve
by a top-hat transfer function with a certain lag,
and fits the model to the data of flux variations in different bands,
one as the source light curve and the others as the response light curves.
The source light curve is modeled as a stochastic process
using a DRW model,
and the posterior distributions of the time lag
as well as other model parameters are estimated using
the Bayesian Markov Chain Monte Carlo (MCMC) method.

%%%%%%%%%%%table 5%%%%%%%%%%
\begin{deluxetable}{ccccccc}[t]
\tablewidth{0pt}
\tablecaption{Time lag of the X-ray flux variation behind that of the $B$ band}
\tablehead{
  \colhead{analysis} & \colhead{$\sigma_{B, \rm sys}$\tablenotemark{a}} &
  \multicolumn{5}{c}{lag time \tablenotemark{b}} \\
  \colhead{}        & \colhead{(mJy)} &
  \multicolumn{5}{c}{(day)} \\
  & & \colhead{1\%} & \colhead{15.9\%} & \colhead{50\%} & \colhead{84.1\%} & \colhead{99\%} }
\startdata
$\tau_{\rm cent}$\tablenotemark{c}
 & $0.00$ & $-3.14$ & $-2.56$ & $-2.09$ & $-1.54$ & $-0.73$ \\
 & $0.06$ & $-3.25$ & $-2.49$ & $-1.95$ & $-1.41$ & $-0.64$ \\
$\tau_{\rm JAV}$\tablenotemark{d}
 & $0.00$ & $-4.42$ & $-2.97$ & $-1.90$ & $-1.32$ & $-0.66$ \\
 & $0.06$ & $-4.07$ & $-2.95$ & $-2.12$ & $-1.53$ & $-0.90$
\enddata
\label{tab:lag}
%\tablecomments{}
\tablenotetext{a}{The systematic error for the $B$-band photometries added to the flux errors in quadrature.}
\tablenotetext{b}{Percentiles.}
\tablenotetext{c}{The time lag based on the CCF analysis.}
\tablenotetext{d}{The time lag based on the {\tt JAVELIN} software.\\\\}
\end{deluxetable}

%%%%%%%%%%%table 5%%%%%%%%%%

% \bibitem[Zu et al.(2011)]{2011ApJ...735...80Z} Zu, Y., Kochanek, C.~S., 
% \& Peterson, B.~M.\ 2011, \apj, 735, 80 
% \bibitem[Grier et al.(2012)]{2012ApJ...755...60G} Grier, C.~J., Peterson, 
% B.~M., Pogge, R.~W., et al.\ 2012, \apj, 755, 60 
% \bibitem[Shappee et al.(2014)]{2014ApJ...788...48S} Shappee, B.~J., Prieto, 
% J.~L., Grupe, D., et al.\ 2014, \apj, 788, 48 
% \bibitem[McHardy et al.(2014)]{2014MNRAS.444.1469M} McHardy, I.~M., 
% Cameron, D.~T., Dwelly, T., et al.\ 2014, \mnras, 444, 1469 
%\bibitem[Koshida et al.(2014)]{2014ApJ...788..159K} Koshida, S., Minezaki, 
%T., Yoshii, Y., et al.\ 2014, \apj, 788, 159 

We use the $B$-band light curve as the source, 
whereas the X-ray data as the response, again with $\tau > 0$ meaning X-ray delays. 
The X-ray data of epoch 1--5 
and the optical data of MJD $=56316$--$56511$
were selected for the {\tt JAVELIN} analysis,
and $\sigma_{\rm add} = 0$ mJy or 0.06 mJy was added to the $B$-band photometric errors
in quadrature, just as in the CCF analysis. 
The posterior distribution of the time lag
calculated by 250000 realizations of the {\tt JAVELIN} software
are presented in Figure \ref{fig:javelin}, and the resultant $\tau_{\rm JAV}$
and its uncertainties are also listed in Table \ref{tab:lag}.
% 0.00 :   -1.904 +0.582 -1.068;  99% -0.660
% 0.06 :   -2.121 +0.588 -0.833;  99% -0.904
% average: -2.0125+0.5925-0.9505; 99% -0.72
The two values of $\sigma_{\rm add}$ again gave very similar results. 
The lag times were estimated as 
$\tau_{\rm JAV}=-1.90^{+0.58}_{-1.07}$ days
and
$\tau_{\rm JAV}=-2.12^{+0.59}_{-0.83}$ days
(1$\sigma $ error) with $\sigma_{\rm add} = 0$ and 0.06, respectively. 
Again, we can exclude the case of $\tau = 0$, 
because the $99$\%-confidence limit is 
$\tau < -0.78$ days on average. 
In summary, the two methods have given consistent results.

\subsection{The X-ray Reflection Component}

As shown in Table \ref{tab:specfit} and Figure \ref{fig:XOLC} (top),
we  found a significant flux variation in the X-ray reflection component 
as well, on a time scale of several months, 
although its variation amplitude is much smaller than that of 
the primary X-rays. 
%  99.9% significant for all epochs from kai^2 fitting of flux=constant
% <97.5% significant for first 5 epochs from kai^2 fitting of flux=constant
This indicates that the source region of the reflection component has
an extent on a scale of $\sim 0.1$ pc.
Moreover, the flux variation of the reflection component slightly lags 
behind that of the hard PL component, on a scale of a week,
% time difference between epoch 4 and 5
or possibly larger because
the X-ray sampling becomes very sparse after epoch 5.
% for 1st 5 epochs, the data can be overfitted even by an equation of flux= a*time + b,
% which means peak location cannot be identified; probable between epoch 4,5 and 6.

According to the unified model of AGNs,
the reflection component accompanied by the neutral Fe-K$\alpha$ line
is supposed to arise from a dust torus, but different origins
such as outer accretion disks, and the broad emission-line region
are also suggested 
(e.g. Awaki et al. 1991; Yaqoob \& Padmanabhan 2004; Nandra 2006;
 Jian et al. 2011; Minezaki \& Matsushita 2015; Gandhi et al. 2015).
% too many?
Interestingly, the reverberation studies of NGC 3516 yielded 
the dust lag of $\sim 50-70$ days (Koshida et al. 2014),
and the lag of broad Balmer emission lines of $\sim 7-13$ days 
(Peterson et al. 2004; Denney et al. 2010).
They are comparable to the time scale of the variation of the X-ray reflection, 
and its delay behind the hard PL component.  
In order to further examine the origin of the X-ray reflection component,
a direct comparison of its variation with that of the dust emission
and the broad Balmer emission lines would be  necessary; 
this will be discussed in a forthcoming paper.

\vspace{0.3cm}
%=============================
\section{Discussion and Conclusion}
%=============================

%--------------------------------------------------
\subsection{Summary of the results}
%--------------------------------------------------

In the present study, we performed an X-ray--optical simultaneous monitoring 
of the type 1.5 Seyfert galaxy NGC 3516 with \textit{Suzaku} and Japanese 
ground-based telescopes, the Pirka, Kiso schmidt, MITSuME, Nayuta, and Kanata telescopes. 
By applying the spectral fitting to the X-ray data and differential image photometry 
to the $B$-band images, and quantitatively comparing the X-ray and optical flux variations, 
we have obtained the following results. 
\begin{itemize}

\item During our observations, NGC 3516 was in an X-ray-faint state. 
It was faintest in epoch 7, when the 2--10 keV flux became 
$\sim 0.21 \times 10^{-11}$ erg s$^{-1}$ cm$^{-2}$, 
which is just  5\% of the average flux recorded in 1997--2002 with \textit{RXTE}.  
Even when brightest (epoch 4), the 2--10 keV flux was $\sim 2.70 \times 10^{-11}$ erg s$^{-1}$ cm$^{-2}$, 
which is only 70\% of the average in 1997--2002.
The flux varied on time scales longer than $\sim$days, without any intraday changes. 

\item The 2--45 keV emission mainly consisted of two spectral components;  
a variable hard power-law continuum with a photon index of $\sim 1.75$, 
and a reflection component with a prominent narrow Fe-K$\alpha$ emission line. 
Throughout the monitoring, the hard X-ray component kept almost the same spectral 
shape, and exhibited a peak-to-peak flux change by $\sim 2.5 \times 10^{-11}$ erg s$^{-1}$ cm$^{-2}$, 
or by about an order of magnitude. 

\item The $B$-band flux density varied by $\sim$ 4.0 mJy in peak to peak, 
which translates to a flux change by
$\sim 2.7\times 10^{-11}$ erg s$^{-1}$ cm$^{-2}$ in $\nu F_{\nu}$ units,
after correcting for the optical extinction. 
The $B$-band flux varied on time scales longer than $\sim$days,    
like those of the X-rays. 

\item X-ray and $B$-band flux correlation was significantly detected 
for the first time in NGC 3516. 
The flux changes of the hard X-ray component significantly preceded 
those in $B$-band by $2.0^{+0.7}_{-0.6}$ days ($1\sigma$ error). 
%After correcting for this lag time, the $B$-band flux showed almost 
%linear correlation to that of the hard power-law continuum. 

\end{itemize}

%--------------------------------------------------
\subsection{X-ray--optical correlation appearing only in faint state}
%--------------------------------------------------

By the previous monitoring of NGC 3516 conducted 
with \textit{RXTE} and the Israeli \textit{WISE} telescope in 1997--2002, 
no significant correlations were derived between 
X-ray and $B$-band flux variations, giving a cross correlation coefficient of $<0.35$ (Maoz et al. 2002). 
In contrast, we succeeded in detecting a significant correlation between them with 
 a high cross correlation coefficient of $>0.95$. 
What is responsible for the clear difference? 
Maoz et al. (2002) argued that the lack of correlation might be due to 
absorption changes that are independent of the primary X-ray variations. 
However, the 2--10 keV flux varied during their monitoring  
by a factor of $\sim 4$ in peak to peak. 
Such a large change would be hardly explained by the so-far recored variations 
in the column density of optically-thick neutral absorbers (Turner et al. 2011). 
We hence suggest alternatively that the dominant 
X-ray variable component changed between the two monitoring campaigns. 

Noda et al. (2013) discovered that the X-ray emission of NGC 3516 comprises  
at least two different primary continua with distinct spectral shape and flux variability; 
a flat ($\Gamma \sim 1.1$--1.7) and steep ($\Gamma \sim 2.3$) spectral continua, 
which we hereafter call the Hard and Gradually-varying Primary Component (HGPC), 
and the Soft and Rapidly-varying Primary Component (SRPC), respectively. 
Noda et al. (2013) showed that luminosities of the HGPC and SRPC 
were comparable in a bright state in 2005, 
when the absorption-corrected 2--10 keV total flux was
$F_{\rm total} \sim 3.5 \times 10^{-11}$ erg s$^{-1}$ cm$^{-2}$.
In contrast, the 2009 Suzaku observation caught NGC 3516 in a faint state
with $F_{\rm total} \sim 1.1 \times 10^{-11}$ erg s$^{-1}$ cm$^{-2}$,
wherein the variation was carried solely by a $\Gamma=1.7$ PL
which can be identified with the HGPC.
Thus, there is a certain threshold in between these two flux values,
say, at $F_{\rm th} \sim 3 \times 10^{-11}$ erg s$^{-1}$ cm$^{-2}$;
above $F_{\rm th}$ the HGPC and SRPC coexist,
whereas only the HGPC remains below $F_{\rm th}$.
These two primary continua have also been identified
in another Seyfert NGC 3227 (Noda et al. 2014): 
\footnote{The HGPC and the SRPC correspond to the Faint-branch Variable (FV)
    and the Bright-branch Variable (BV) components in Noda et al. (2014),
    respectively. }
Their  luminosity-dependent behavior was
found to be very similar to that observed from NGC 3516.
Namely, the HGPC was always present,
whereas the SRPC appeared when the source 
was brighter than a certain threshold,
to become co-existent with the other.

In the present \textit{Suzaku} observations, the absorption-corrected 2--10 keV flux of NGC 3516
was in the range $(0.2-2.7) \times 10^{-11}$ erg  s$^{-1}$ cm$^{-2}$, i.e., 
always below the suggested $F_{\rm th}$. 
Furthermore, the 2--45 keV spectrum 
of the main variable PL was flat with $\Gamma \sim 1.75$, 
and the CCP distributions are smoothly connected to those in 2009 (Figure \ref{fig:Xccp}).  
Hence, we conclude that NGC 3516 was in the faint phase, 
with the HGPC dominating just as in the 2009 observation. 
On the other hand, during the 1997--2002 monitoring by Maoz et al. (2002),  
the averaged 2--10 keV flux was $F_{\rm total} \sim 4.0 \times 10^{-11}$ erg s$^{-1}$ cm$^{-2}$ 
which is higher than $F_{\rm th}$, 
indicating that NGC 3516 was mostly in the bright phase. 
If the SRPC has a significantly poorer correlation with the optical than the HGPC does,  
the correlation between the total X-ray flux and the optical should become worse  
when NGC 3516 is in the bright phase. 
Thus, the difference between Maoz et al. (2002) and ours 
may be explained by presuming that the HGPC is well correlated with the optical 
while the SRPC is not. 

To generalize the above conclusion, 
we collected cross correlation coefficients (CCCs) from peak values 
of the CCF in previous simultaneous observations of various Seyferts. 
As a result, sources with high $L_{\rm bol}/L_{\rm Edd}$ ratios were found to have relatively low CCCs;
MCG--6-30-15 with $L_{\rm bol}/L_{\rm Edd} \sim 0.2$ has $\textrm{CCC} \lesssim 0.1$ (Lira et al. 2015),   
NGC 4051 with $L_{\rm bol}/L_{\rm Edd} \sim 0.15$ has $\textrm{CCC} \sim 0.3$ (Breedt et al. 2009), 
and MR 2251-178 with $L_{\rm bol}/L_{\rm Edd} \sim 0.2$ has $\textrm{CCC} \sim 0.5$ (Arevalo et al. 2008). 
In contrast, sources with low  $L_{\rm bol}/L_{\rm Edd}$ ratios possibly have high CCCs; 
NGC 6814 with $L_{\rm bol}/L_{\rm Edd} \sim 0.008$ has $\textrm{CCC} \sim 0.9$ (Troyer et al. 2015); 
and NGC 3516 in the present study has $\textrm{CCC} \gtrsim 0.95$ 
wherein  $L_{\rm bol}/L_{\rm Edd} < 0.01$ is indicated by 
the value of  0.00612 (Vasudevan \& Fabian 2009), obtained in 2001 
when NGC 3516 showed $F_{\rm tot}$ close to the second highest among ours. 
%and NGC 3516 $L_{\rm bol}/L_{\rm Edd} \lesssim 0.01$ has $\textrm{CCC} \gtrsim 0.95$ in the present study. 
Therefore, we speculate that X-ray--optical correlation in Seyfert galaxies
becomes commonly worse when the Eddington ratio gets higher, 
because the SRPC start dominating. 

According to Figure 12(b) and related descriptions in Noda et al. (2014), 
the SRPC may originate from patchy coronae heated by magnetic process on the surface 
of a disk (e.g., Reynolds \& Nowak 2003).  
If covering fraction of the patchy coronae against the disk is small, 
the SRPC generated in the coronae can be related to just small areas of the disk, 
making X-ray and optical variations independent. 
Another possible origin of the SRPC 
is a faulty jet at the accretion axis of the SMBH (e.g., Ghisellini et al. 2004). 
If the jet emission region is located far away from the accretion disk 
as suggested by Noda et al. (2014), connection between the SRPC and 
the disk optical emission may become weaker, making their correlation poorer. 
In any case, the emergence and disappearance of these SRPC-generating regions 
are likely to be more localized effects than those responsible for the hard vs. soft state 
transitions seen in black-hole binaries, 
because the AGN state changes considered here 
take place on time scales of weeks to years, 
which are much shorter than would be expected (e.g., several ten thousand years) 
if the typical state-transition time scales, $\sim$days, are scaled to the mass ratios.

%--------------------------------------------------
\subsection{The origin of X-ray and optical correlation}
%--------------------------------------------------

%--------------------------------------------
\subsubsection{Application of standard X-ray reprocess model}
%--------------------------------------------

The measured optical lag of $2.0^{+0.7}_{-0.6}$ days strongly supports 
the X-ray reprocessing model (e.g., Krolik et al. 1991; Cackett et al. 2007); 
optical brightening occurred, at least in the present case, 
through irradiation by the increased hard X-ray intensity.  
This process is energetically feasible, 
because the 2--10 keV X-ray flux increase by $2.5\times10^{-11}$ erg s$^{-1}$ cm$^{-2}$
is comparable to the $B$-band flux increase by 
$\sim 2.7\times10^{-11}$ erg s$^{-1}$ cm$^{-2}$. 
The optical lag time in NGC 3516 obtained here is similar to 
those of a few days observed from other Seyfert galaxies, 
including NGC 5548  (e.g., Suganuma et al. 2006; Mchardy et al. 2014; Edelson et al. 2015), 
NGC 3783 (Arevalo et al. 2009), NGC 4051 (Breedt et al. 2010), and NGC 6418 (Troyer et al. 2015).
Therefore, the suggested mechanism, i.e., the X-ray reprocessing, 
can be operating commonly among these Seyfert galaxies.

Let us, then, investigate whether or not the derived optical lag 
can be explained by the most commonly adopted X-ray reprocessing model (e.g., Cackett et al. 2007),
which assumes that an optically-thick  geometrically-thin standard accretion disk 
continues down to the innermost last-stable circular orbit (ISCO),
located at $3R_{\rm s}$ 
where $R_{\rm s} = 2GM_{\rm BH}/c^2$ is the Schwarzschild radius
and $M_{\rm BH}$ the black-hole mass,
and is illuminated by a lamppost-type X-ray source proximate to the black hole.
At radii $R \gg R_{\rm s}$,
the radial temperature profile of the disk is given as (Shakura \& Sunyaev 1973)
\begin{equation}
T(R) = \left(\frac{3GM_{\rm BH}\dot{M}}{8\pi \sigma}\right)^{\frac{1}{4}}R^{-\frac{3}{4}}\ \ \ ,
\end{equation}
where $\dot{M}$ is the mass accretion rate,
$G$ is the gravitational constant,
and $\sigma$ is the Stefan-Boltzmann constant.

%Newtonian constant of gravitation,
%When the radius is much greater than the inner radius of the accretion disk,
%the viscous dissipation per unit disk face area is given by

After the delay time $\tau$,
the X-ray flux variation that arose close to the black hole 
will propagate to a radius $R=c \tau$,
and will increase the emissivity there.
Let us define a wavelength $\lambda =hc/ kT(c \tau) X$
in such a way that the continuum emission 
at this radius is peaked at $\lambda$,
where $X$ is a numerical factor of order unity.
Although the Wien's displacement law for  a blackbody radiation
in $\nu B_{\nu}$ gives $X=3.92$,
we here set $X\sim 3.2$,
according to the detailed calculations
of the disk response to  continuum emission
by Cackett et al. (2007) and  Collier et al. (1999).
Substituting $c \tau $ for $R$,
and expressing $T(c\tau)$ with  $\lambda$,
eq. (1)  can be rewritten as
\begin{equation}
T(R)=\frac{hc}{k\lambda X}\left(\frac{R}{c\tau }\right)^{-\frac{3}{4}}\ \ \ .
\end{equation}
By eliminating $T(R) \cdot R^{3/4}$ from eq. (1) and eq. (2),
we obtain $\dot{M} \propto M^{-1} \tau^3 \lambda^{-4}$.
Further expressing  $\dot{M}$ by the 
bolometric luminosity of the accretion disk, 
$L_{\rm bol} \sim 0.1\ \dot{M}c^2$ (assuming that the disk extends to the ISCO),
and normalizing it to the Eddington luminosity ($\propto M_{\rm BH}$),
the Eddington ratio of the source $\eta$ is estimated as
\begin{eqnarray}
\eta \equiv L_{\rm bol}/L_{\rm Edd} \sim  
3.8\ \left(\frac{M_{\rm BH}}{3.2\times 10^7\ M_{\odot}}\right)^{-2} \nonumber 
\\ 
\times \left(\frac{\tau}{2.0\ {\rm day}}\right)^3 \left(\frac{\lambda}{0.44\ \mu {\rm m}}\right)^{-4}\ \ \ .
\end{eqnarray}

Equation (3), in fact, implies a serious problem.
First of all, it indicates an unrealistic ``super-Eddington" condition.
Even putting aside this  issue,
it means a factor of $\sim 500$ discrepancy 
against the likely value of $\eta=0.006- 0.01$ (\S~4.2),
which characterizes the NGC 3516 nucleus during the present observations.
The difficulty can be more directly stated:
by summing up the blackbody contributions from various disk annuli,
the continuum spectrum from the accretion disk under consideration can be calculated as
\begin{eqnarray}
f_{\nu }\sim 550\ {\rm mJy}\ \left(\frac{\tau}{2.0\ {\rm days}}\right)^{2}\left(\frac{D}{41.3\ {\rm Mpc}}\right)^{-2} \nonumber \\ 
\times \left(\frac{\lambda}{0.44\ \mu {\rm m}}\right)^{-3} \cos i\ \ \ ,
\end{eqnarray}
where $i$ is the disk inclination (Collier et al. 1999; Cackett et al. 2007).
Thus, the measured delay predicts very bright optical emission,
arising from the inner part of the accretion disk.
In other words, the SMBH would have to be accreting with a very high rate,
in order for its accretion disk to emit the $B$-band light
from such a large distance as $\sim 2$ light days.
Equation (4) means a two-orders-of-magnitude contradiction to the present optical data,
that the peak-to-peak flux variation in the $B$ band was $\sim 4.0$ mJy,
and the minimum $B$-band flux during the observation would be
less than that.
Inversely, if we started from the assumption of
$\eta = 0.006$ -- $0.01$,
the $B$-band lag predicted by eq.(3) would become
$\sim 8$ times smaller than was measured.
It is thus extremely difficult to reconcile
the measured clear optical delay by $\sim 2$ days
with the observed optical faintness,
as long as we assume a standard disk extending down to the ISCO
and an illuminating X-ray source close to the SMBH.

%--------------------------------------------
\subsubsection{Consideration of the X-ray irradiation}
%--------------------------------------------

In the X-ray reprocessing model, 
we should consider not only the viscous heating but also the heating by the X-ray irradiation.
The former, which underlies eq.(1), is written as
\begin{equation}
D_{\rm vis} = \frac{3GM_{\rm BH}\dot{M}}{8\pi R^3}\ \ \ ,
\end{equation}
whereas the latter is given as
\begin{equation}
 D_{\rm irr} = \frac{(1-A)L_{\rm X}}{4\pi R_{\rm X}^2}\cos \theta \ \ \ ,
\end{equation}
where $L_{\rm X}$ is the X-ray luminosity, $A$ is the disk albedo, 
$R_{\rm X}$ is the distance from the disk to the X-ray illuminator, 
and $\theta$ is the X-ray incidence angle onto the disk.
Collier et al. (1999) and Cackett et al. (2007) assumed that 
the X-ray emitter is located on the rotating axis of the  SMBH like a ``lamppost'' 
(Figure \ref{fig:model}a), 
and its height $H_{\rm X}$ from the SMBH
is much smaller than $R$ ($H_{\rm X} \ll R$).
Under these assumptions, we obtain $R_{\rm X} \sim R$ 
and $\cos \theta = H_{\rm X}/\sqrt{R^2+H_{\rm X}^2} \sim H_{\rm X}/R$. 
Then, $ D_{\rm vis}$ and  $D_{\rm irr}$ have just the same $R$-profile,
and the total heating per unit disk face area,
$D_{\rm tot} = D_{\rm vis} +  D_{\rm irr}$, is described as
\begin{equation}
D_{\rm tot} = \frac{3GM_{\rm BH}\dot{M}}{8\pi R^3} 
 + \frac{(1-A)L_{\rm X}H_{\rm X}}{4\pi R^3} ~~.
\end{equation}
As a result, the radial temperature profile of the disk is now given by
\begin{equation}
T(R)= \left[ \frac{3GM_{\rm BH}}{8\pi \sigma}
 \left\{ \dot{M}+\frac{2(1-A)L_{\rm X}H_{\rm X}}{3GM_{\rm BH}}\right\} 
 \right]^{\frac{1}{4}}R^{-\frac{3}{4}}.
\end{equation}
It exhibits the same $R$-dependence as  the standard viscous disk, 
namely, $T\propto R^{-3/4}$.
%by just substituting $\dot{M}+2(1-A)L_{\rm X}H_{\rm X}/(3GM_{\rm BH})$ for $\dot{M}$
%in equation (2),
%
Furthermore, because of the $L_{\rm X} \propto \dot{M}$ proportionality,
we should use
 $\dot{M}+ 2(1-A)L_{\rm X}H_{\rm X}/3GM_{\rm BH}$ in place of $\dot{M}$
when converting this expression into that of $\eta$ as eq.(3).
Consequently, the Eddington ratio $\eta$ and the $B$-band flux density can be given by 
the same eq.(3) and eq.(4), respectively.
\footnote{
The Eddington ratio would differ by a factor of order unity
because the radial profile of the temperature of the irradiated disk
would be somewhat different from that of the standard viscous disk at small radii.}
Thus,  we encounter just the same problem as before,
even when the X-ray irradiation is taken into account.

According to e.g., Cackett et al. (2007), McHardy et al. (2014), and Edelson et al. (2014), 
typical Seyfert galaxies including NGC 5548, NGC 4051, and Mrk 335 
yielded $\tau$ that are a few times larger than 
those expected from  their estimated $\eta$.
According to Mogan et al. (2010),  
the accretion disk sizes  of gravitationally lensed quasars are  larger,
by a factor of $\sim 4$, than those expected from $\eta$, 
just like in the Seyfert galaxies.
These subtle but systematic contradictions 
between the disk size and $\eta$ have also
been well known in the X-ray reprocessing scenario 
(e.g., Collier et al. 1999, Cackett et al. 2007). 
Recently, Troyer et al. (2015) reported that NGC 6814 exhibited $\tau$ 
which is $\sim40$ times larger than that predicted from the estimated $\eta$. 
In the present case of NGC~3516,
the measured vs. predicted difference in the optical delay 
amounts to a factor of $\sim 10$,
which is considerably larger than those of NGC 5548, NGC 4051, and Mrk 335.
These systematic discrepancies imply 
that the lamppost-type X-ray irradiation geometry shown in Figure \ref{fig:model}(a)
has a common problem, 
which is considered to become more prominent towards lower luminosities,
considering that NGC 3516 and NGC 6814 have both relatively low values of $\eta$.
Thus, we need to revise  the geometrical assumptions,
in order to make $\tau$ and $\eta$ consistent in low Eddington-ratio Seyfert galaxies. 

%--------------------------------------------------
\subsubsection{Possible geometry of accretion flows in low Eddington-ratio Seyferts}
%--------------------------------------------------

%%%%%%%%%%%%%%%%%%figure11%%%%%%%%%%%%%%%%%%%
\begin{figure*}[t]
\epsscale{1}
\plotone{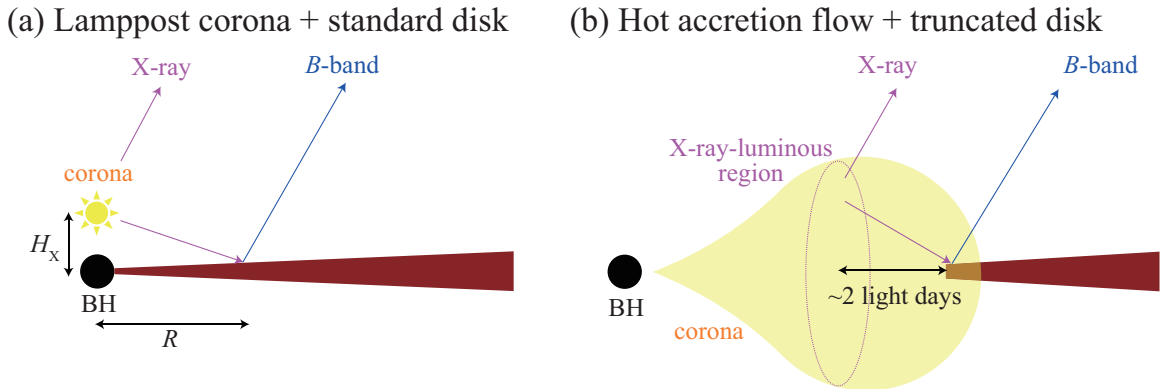}
\caption{Schematic pictures of geometries of an accretion disk and a corona,
in the lamppost configuration (panel a) 
and the model invoking a truncated disk and a hot accretion flow inside (panel b). \\\\}
\label{fig:model}
\end{figure*}
%%%%%%%%%%%%%%%%%%figure11%%%%%%%%%%%%%%%%%%%%

What kind of geometry of the corona and the accretion disk can  
reconcile  $\eta$ and $\tau$ from the present study?
The contradiction in the lamppost-type X-ray reprocessing model,
revealed in \S4.3.1, arises mainly because
the observed luminosity is too low for the size of the accretion disk
determined by the observed time lag.
Conversely, the measured $\tau\sim 2$ days
is too large to be explained by the standard scenario,
given the very low luminosity. 
In this respect, it has been very essential
that the present observations caught the object 
at a very faint state 
not only
in X-rays but also in the optical band (see the final paragraph in \S3.2).
This indicates that the $\eta - \tau$ inconsistency cannot be 
solved even if considering an anisotropically-irradiating lamppost 
corona (Dabrowski et al. 1997)
that would make the X-ray flux we observe much lower than that 
illuminating accretion disk. 
Hence, we need to consider other geometries of the corona and the accretion disk 
that can reconcile  $\eta$ and $\tau$ from the present study.

One possible variant of the lamppost-type scenario is
to place the X-ray source at a large height from the SMBH,
to achieve $H_{\rm X} \sim 2$ light days.
In this geometry, X-ray flux variations can still precede that in the optical. 
Furthermore, the inner part of the accretion disk  can be made rather cool,
because $\dot{M}$ can be reduced and $F_{\rm irr}$ gets much smaller 
than that with the assumption of $H_{\rm X} \ll R$. 
These would make the black body flux low enough
to be consistent with the observed optical faintness.
Such a  lamppost-type corona with $H_{\rm X} \lesssim  R$ 
was suggested to form 
at the base of a jet moving away from a SMBH with relativistic speeds 
(e.g., Lohfink et al. 2013;  Wilkins \& Gallo 2015).
However, the value of $H_{\rm X} \sim 2$ light days,
or $\sim 500~R_{\rm s}$,
which is required to explain the present results, 
is  orders of magnitude larger than invoked in the previous studies.
It would be highly unrealistic to assume that such an emission region 
forms at a height of  $\sim 500~R_{\rm s}$ in this radio-quiet object,
and that the region radiates essentially the {\it entire primary X-ray emission}  (\S 3.1).
In addition, the X-ray emission mechanism from such a region is problematic.
As observed from BL-Lac objects,
synchrotron radiation would give an X-ray spectral slope of $\Gamma \sim 2.5$  
(e.g., Tanihata et al. 2004), %<=== http://ads.nao.ac.jp/abs/2004ApJ...601..759T
which is much steeper than we observed ($\Gamma \sim 1.75$).
Comptonization of soft photons from the accretion disk
would  produce {\it X-ray delays},
contrary to the observed optical delay.
In short, this variant scenario fails to explain the present results.

As we noted repeatedly, 
the X-ray reprocessing model which has so far been considered  (Figure \ref{fig:model}a)
assumes the optically-thick accretion disk to extend down to near the ISCO.
However, it has long been known theoretically  
(e.g., Abramowicz et al. 1995) %<== http://ads.nao.ac.jp/abs/1995ApJ...438L..37A
that a decreased luminosity,  e.g., to $\eta \lesssim 0.01$,
causes  inner regions of such a disk to make a transition 
into an optically-thin and geometrically-thick hot flow.
This will lead to a configuration  shown in Figure \ref{fig:model}(b);
the disk is truncated at a radius much larger than the ISCO,
and the accretion flow inside that radius may  form
a  Radiatively Inefficient Accretion Flow, or RIAF (e.g., Yuan \& Narayan 2014).
Such bimodal behavior of accretion disks has been 
firmly verified through extensive observations of black-hole binaries
(e.g., Done et al. 2007).
The truncated disk picture has also  been investigated 
by many authors, both observationally and theoretically,
to explain low-luminosity AGNs
(e.g. Ho 2008; Taam et al. 2012; Nemmen et al. 2014; and references therein).
It is hence much more natural to try to explain
the present results based on the geometry of Figure \ref{fig:model}(b).

Referring to Figure \ref{fig:model}(b),
let us then consider an accretion disk 
that is truncated at a radius $R\gtrsim 2$ light days 
(or $\sim 500$ times there Schwarzschild radii) from the SMBH.
Let us also assume that  the accreting matter inside this radius  forms a RIAF region, 
where the illuminating hard X-ray photons are produced via Comptonization
(Noda et al. 2013a, 2014; Makishima et al. 2008; Yamada et al. 2013).
In such a hot RIAF region,
the electron density and temperature are theoretically 
predicted to increase inwards towards the SMBH (e.g., Esin et al. 1997),  
whereas the seed photon flux for inverse Comptonization 
obviously increases towards positions closer to the disk.  
Hence,  the X-ray emissivity are considered to become maximum
at a location in between the SMBH and the inner disk edge.
Then, the optical delay of $\tau \sim 2$ days can be explained
simply as the light travel time from the X-ray brightest regions 
to the inner disk edge.
Of course, this scenario requires a relatively low inclination,
in order not to produce large differences in the light travel time to us,
from the near side and far side of the disk;
NGC 3516 may satisfy the condition,
because of a relatively low inclination value 
of $i \sim 38^{\circ}$ (Wu \& Han et al. 2001).

The truncated-disk picture can naturally explain,
at the same time, the optical faintness,
because the large amount of continuum radiation originating 
in the inner part of the accretion disk is no longer present.
The $B$-band to X-ray flux ratio of NGC 3516 in these epochs
is $\nu F_{\nu}(B)/F_{X}(2-10\ {\rm keV}) \approx 1$.
It is much lower than that of the composite spectrum
for radio-quiet quasars ($\sim 6$; Shang et al. 2011),
but is consistent with those of the composite and model spectra
for low-luminosity AGNs (Ho 2008; Nemmen et al. 2014).
To be quantitative, let us assume, in eq. (6), 
$L_{\rm X} = 5.2\times10^{42}$ erg s$^{-1}$, 
$A=0$, $R_{\rm X} = 2$ light days,  and $\theta=\pi/4$.
Then, the illuminating X-ray flux at a distance of 2.0 light days 
from the X-ray-bright region is calculated as 
$D_{\rm irr} \sim 1.6 \times 10^{10}$ erg s$^{-1}$ cm$^{-2}$ 
during the peak of the flux variation. 
It predicts the disk temperature at the inner radius to be $\sim 4000$K 
wherein $B$-band photons can be produced on the Wien side, 
even when the viscous heating is ignored in eq. (8). 
Incidentally, the disk temperature would increase
by a factor of  $(3/500)^{-3/4} \sim 45$
if the disk extended down to the ISCO.

As another merit of the truncated-disk scenario,
the gradual X-ray variations without any intraday changes 
can be explained as well by the large volume of the RIAF region. 
In addition, the observed  hard X-ray spectrum  with $\Gamma \sim 1.7$ can be
naturally explained as arising from the hot RIAF region via thermal Comptonization,
as theoretically predicted (Esin et al. 1997),
and observed from black-hole binaries (e.g., Makishima et al. 2008).
Moreover, as shown in Figure \ref{fig:LowEneSpec}, 
the soft X-ray band at epoch 7 was dominated by the galactic thin-thermal 
plasma emission, and no relativistically-smeared spectral components were required, 
also supporting that the standard accretion disk did not continue down to the ISCO.
This scenario may be able to solve the $\tau$ vs. $\eta$ conflict
in  the other low-$\eta$ Seyferts as well,
including NGC 6814 in particular (Troyer et al. 2015). 
We hence  prefer the picture of the truncated disk and RIAF 
shown in Figure \ref{fig:model}(b).

How does our model compare with other attempts on the same subject? 
Recently, Gardner \& Done (2016) proposed a model, that a geometrically-thick region, 
called the soft-excess region, is present at the inner edge of an accretion disk,  
and completely hides the hard X-ray corona. 
Far-ultraviolet photons, arising from the soft-excess region by the X-ray illumination,  
are reprocessed by optically-thick clouds distributing above the accretion disk, 
and produce optical variations. 
This model is different from ours in that X-rays from the corona cannot directly 
reach the accretion disk, and the optical variation is mainly due to
the clouds rather than the disk. 
Presumably, our NGC 3516 results cannot be explained by the model of Gardner \& Done (2016), 
because the 0.5--2 keV spectrum at epoch 7 is explained adequately by the galactic emission,  
without any strong soft excess resulting from the AGN activity. 
Therefore, we prefer our model (Figure \ref{fig:model}b) to theirs.

For some quasars, it is claimed that the accretion disks, 
as measured by the UV-optical microlensing observations, are generally
several times larger than those predicted by the standard 
accretion disk model (e.g., Morgan et al. 2010). 
These measurements  are thought to be observing the unexpectedly large disk size 
like those revealed by the X-ray-to-optical continuum reverberation mapping for
nearby Seyfert galaxies (e.g., Edelson et al. 2015, McHardy et al. 2016). 
Dexter \& Agol (2011) suggested that the large disk size of 
quasars from microlensing is due to stochastic and strong local 
temperature fluctuations on the effective temperature profile of 
the standard accretion disk.
Their inhomogeneous disk model is designed to enhance the flux
contribution from outer disk radii, and thus make the disk 
half-light radius larger than the predictions of the standard accretion disk. 
However, their model, which does not consider the X-ray emission and attributes 
all UV--optical variability to the assumed temperature fluctuations, cannot explain 
the strong X-ray vs. optical correlations observed in NGC 3516 and other Seyfert
galaxies. 
Furthermore, Kokubo (2015) showed that the disk model in Dexter \& Agol (2011) 
cannot explain, either, the tight inter-band correlations often observed in
the optical light curves of quasars. 
Therefore, we do not discuss about the model any further.

For further refinement of the truncated-disk model,
 and its demarcation from the large-$H_{\rm X}$ model, 
correlations and time lags among different optical bands, 
including $U$, $V$, and $R$ bands become important 
(e.g., Kokubo et al. 2014; Edelson et al. 2015; Fausnaugh et al. 2015; Kokubo 2015). 
These will be discussed in elsewhere.

\subsection{Conclusion}
The present coordinated X-ray and optical observations of NGC 3516
found the object in a very faint state in both wavelengths.
A tightly correlated intensity change lasting for  $\sim 60$ days was detected, 
wherein the optical $B$-band variation showed a clear delay 
by $\sim 2$ days behind that of the X-ray continuum
which is described by a power-law model with $\Gamma \sim 1.7$.
This optical lag indicates the effect of X-ray reprocessing,
but the delay cannot be reconciled with the $B$-band faintness,
as long as we consider the standard reprocessing scenario
which invokes a standard accretion disk 
 irradiated by a lamppost-type X-ray source.
Instead, the observed results can be consistently explained
assuming that the disk is truncated at $\sim 500~R_{\rm s}$,
and turns into a RIAF where the illuminating hard X-rays are produced 
via thermal Comptonization.

\section*{Acknowledgement}
We thank the referee for his/her valuable comments and suggestions. 
This work was partly supported by the Grants-in-Aid with grant numbers 26800095 (HN), 
25287062 (TM), and 15J10324 (MK), from the Japan Society for the Promotion of Science (JSPS). 
HN was supported by the Special Postdoctoral Researchers Program in RIKEN.

% figures 

\end{document}